\documentclass[aps,prc,twocolumn,showpacs,amsmath,amssymb,nofootinbib,superscriptaddress]{revtex4-1}
\usepackage{graphicx} 
\usepackage{dcolumn}

\begin{document}
\title{High-spin structures of five $N=82$ isotones: $^{136}_{\ 54}$Xe, 
$^{137}_{\ 55}$Cs, $^{138}_{\ 56}$Ba, $^{139}_{\ 57}$La, and $^{140}_{\ 58}$Ce}
\author{A.~Astier}
\author{M.-G. Porquet}
\affiliation{CSNSM, IN2P3-CNRS and Universit\'e Paris-Sud, B\^at 104-108,
F-91405 Orsay, France}
\author{Ts.~Venkova}
\affiliation{CSNSM, IN2P3-CNRS and Universit\'e Paris-Sud, B\^at 104-108,
F-91405 Orsay, France}
\affiliation{INRNE, BAS, 1784 Sofia, Bulgaria}
\author{D.~Verney}
\affiliation{IPNO,  IN2P3-CNRS and Universit\'e Paris-Sud, F-91406 Orsay, France}
\author{Ch.~Theisen}
\affiliation{CEA, Centre de Saclay, 
IRFU/Service de Physique Nucl\'eaire, F-91191 Gif-sur-Yvette Cedex, France} 
\author{G.~Duch\^ene}
\affiliation{IPHC, IN2P3-CNRS and Universit\'e Louis Pasteur, F-67037 Strasbourg Cedex 2, France}
\author{F.~Azaiez}
\altaffiliation{Present address: IPNO,  IN2P3-CNRS and Universit\'e Paris-Sud, 
F-91406 Orsay, France}
\affiliation{IPHC, IN2P3-CNRS and Universit\'e Louis Pasteur, F-67037 Strasbourg Cedex 2, France}
\author{G.~Barreau}
\affiliation{CENBG, IN2P3-CNRS and Universit\'e Bordeaux I, F-33175 Gradignan, France}
\author{D.~Curien}
\affiliation{IPHC, IN2P3-CNRS and Universit\'e Louis Pasteur, F-67037 Strasbourg Cedex 2, France}
\author{I.~Deloncle}
\affiliation{CSNSM, IN2P3-CNRS and Universit\'e Paris-Sud, B\^at 104-108,
F-91405 Orsay, France}
\author{O.~Dorvaux}
\author{B.J.P.~Gall}
\affiliation{IPHC, IN2P3-CNRS and Universit\'e Louis Pasteur, F-67037 Strasbourg Cedex 2, France}
\author{M.~Houry}
\altaffiliation{Present address: CEA/DSM/D\'epartement de recherches sur la Fusion
Contr\^ol\'ee, F-130108 Saint-Paul lez Durance, France}
\author{R.~Lucas}
\affiliation{CEA, Centre de Saclay, 
IRFU/Service de Physique Nucl\'eaire, F-91191 Gif-sur-Yvette Cedex, France} 
\author{N.~Redon}
\affiliation{IPNL, IN2P3-CNRS and Universit\'e Claude Bernard, F-69622 Villeurbanne Cedex, France} 
\author{M.~Rousseau}
\affiliation{IPHC, IN2P3-CNRS and Universit\'e Louis Pasteur, F-67037 Strasbourg Cedex 2, France}
\author{O.~St\'ezowski}
\affiliation{IPNL, IN2P3-CNRS and Universit\'e Claude Bernard, F-69622 Villeurbanne Cedex, France}

\date{Received: date / Revised version: date}
\date{\hfill \today}

\begin{abstract}
Five $N=82$ isotones have been produced in two fusion-fission reactions and their 
$\gamma$ rays studied with the Euroball array. The high-spin states of $^{139}$La
have been identified for the first time, while the high-spin yrast and near-to-yrast
structures of the four others have been greatly extended. From angular correlation analysis,
spin values have been assigned to some states of $^{136}$Xe and $^{137}$Cs. Several 
cascades involving $\gamma$ rays of $^{139}$La have been found to be delayed, they deexcite
an isomeric state with $T_{1/2}= 315(35)$~ns located at 1800-keV excitation energy.
The excited states of these five $N=82$ isotones are expected to be due to various proton 
excitations involving the three high-$j$ subshells located above the $Z=50$ shell closure.
This is confirmed by the results of shell-model calculations performed in this work. 
In addition, high-spin states corresponding to the excitation of the 
neutron core have been unambiguously identified  in $^{136}$Xe, $^{137}$Cs, and 
$^{138}$Ba.
\end{abstract} 


\maketitle

\section{Introduction}
The study of the low-lying excited states of the $N=82$ isotones offers the best
ground to describe the gradual filling of the active proton orbitals lying above the 
$Z=50$ closure, since these states are free of any neutron contribution up to an 
excitation energy of about 4~MeV. Particularly, the description of the high-spin 
states mainly involve the effective interactions of the protons when occupying the 
valence orbits, particularly the high-$j$ ones, $\pi g_{7/2}$, $\pi d_{5/2}$, and  
$\pi h_{11/2}$, which can thus be tested. 
Many years ago, a lot of low-spin states of several $N=82$ isotones were used to 
define the two-body part of shell-model hamiltonian~\cite{wi91}, which was
adjusted some times later~\cite{bl99} using new data on some high-spin states of 
the ''light'' $N=82$ isotones produced by spontaneous fission 
of actinides~\cite{zh96,da99}. This set of empirical interactions was then tested on
the new high-spin states of $^{137}$Cs, identified in a deep-inelastic reaction up to
I$^\pi$=(31/2$^-$)~\cite{br99}. 

By combining appropriate fusion-fission reactions to a powerful $\gamma$-detection array, 
it is possible to populate and to study the high-spin states of many $N=82$ isotones.  
In this paper,  we present our results obtained on the 
high-spin states of five isotones, $^{136}_{\ 54}$Xe, 
$^{137}_{\ 55}$Cs, $^{138}_{\ 56}$Ba, $^{139}_{\ 57}$La, and $^{140}_{\ 58}$Ce, 
which have been identified in two Euroball experiments. The level schemes of four of
them have been well extended to higher-spin values. The high-spin structures of 
$^{139}$La have been identified for the first time, displaying an isomeric state 
with $T_{1/2}= 315(35)$~ns located at 1800-keV excitation energy. 
The states of these five $N=82$ isotones are then compared to the results of shell-model 
calculations using the set of empirical interactions already used to describe 
the lightest isotones. In addition, groups of states lying above 4~MeV excitation
energy in $^{136}$Xe, $^{137}$Cs, and $^{138}$Ba are assigned to the excitation of the 
neutron core. 
     

\section{Experimental details}
\subsection{Reactions, $\gamma$-ray detection and analysis\label{exp}}
The $N=82$ isotones of interest were obtained as fission fragments in 
two experiments. First, the $^{12}$C + $^{238}$U reaction was studied at 90 MeV incident 
energy, with a beam provided by the Legnaro XTU Tandem accelerator. Secondly, the 
$^{18}$O + $^{208}$Pb reaction was studied with a 85 MeV incident 
energy beam provided by the Vivitron accelerator of IReS (Strasbourg). 
The gamma-rays were detected with the Euroball array~\cite{si97}. 
The spectrometer contained 15 
Cluster germanium detectors placed in the backward hemisphere with 
respect to the beam, 26 Clover germanium detectors located 
around 90$^\circ$ and 30 tapered single-crystal germanium detectors 
located at forward angles. Each Cluster detector consists of seven 
closely packed large volume Ge crystals \cite{eb96} and each 
Clover detector consists of four smaller Ge crystals \cite{du99}.
In order to get rid of the Doppler effect, both experiments 
have been performed with thick targets in order to stop the recoiling nuclei 
(47 mg/cm$^{2}$ for $^{238}$U and 100 mg/cm$^{2}$ for $^{208}$Pb targets, respectively). 

The data of the (C+U) experiment were recorded in an event-by-event mode with the 
requirement that a minimum of five unsuppressed Ge
detectors fired in prompt coincidence. A set of 1.9$\times 
10^{9}$ three- and higher-fold events was available
for a subsequent analysis. For the (O+Pb) experiment, a lower trigger condition 
(three unsuppressed Ge) allowed us to register 4$\times 10^{9}$ events with a 
$\gamma$-fold greater or equal to three. The offline analysis consisted 
of both multi-gated spectra and three-dimensional 'cubes' built 
and analyzed with the Radware package \cite{ra95}.

More than one hundred nuclei are produced at high spin in 
such experiments, and this gives several thousands of $\gamma$ 
transitions which have to be sorted out. Single-gated
spectra are useless in most of the cases. The selection of one 
particular nucleus needs at least two energy conditions, implying 
that at least two transitions have to be known.
The identification of transitions depopulating high-spin 
levels which are completely unknown is based on the
fact that prompt $\gamma$ rays emitted by complementary 
fragments are detected in coincidence \cite{ho91,po96}.
Moreover, the isotopes of interest being produced from two different 
fissionning compound nuclei, the complementary fragments are different in 
the two reactions. This gives a fully unambiguous assignment of transitions 
seen in both experiments.

\subsection{Isomer identification\label{isomer}}
As reported in previous papers~\cite{lu02,po05}, another experiment 
was performed using the SAPhIR\footnote{SAPhIR, Saclay Aquitaine Photovoltaic cells
for Isomer Research.} heavy-ion detector~\cite{Saphir}, here composed 
of 32 photovoltaic cells, in order to identify new isomeric states in the fission 
fragments. Placed in the target chamber of Euroball, SAPhIR was used to 
detect the escaping fission-fragments of the $^{12}$C (90~MeV) + $^{238}$U 
reaction from a thin 0.14 mg/cm$^{2}$ uranium target. Thus, all recorded 
$\gamma$ rays emitted at rest by a fragment implanted in SAPhIR correspond 
to the decay of an isomeric state, for which the half-life can be measured 
in a range of several tens to several hundreds of nanoseconds. 

\subsection{$\gamma$-$\gamma$ angular correlations \label{correl}}
It is well known that the $\gamma$ rays emitted by fusion-fission fragments do not show
any anisotropy in their angular distributions with respect to the incident beam. 
However, angular correlations of two successive transitions are meaningful.
In order to determine the spin values of excited states, the coincidence rates 
of two successive $\gamma$ rays are 
analyzed as a function of $\theta$, the average relative angle between the 
two fired detectors.
The Euroball spectrometer had $C^{2}_{239}$=28441 combinations of 2 crystals, out 
of which only $\sim$ 2000  
involved different values of relative angle within 2$^\circ$. Therefore, in order 
to keep reasonable numbers of counts, all these angles have been 
gathered around three average relative angles : 22$^\circ$, 46$^\circ$, 
and 75$^\circ$. The coincidence rate is increasing between 0$^\circ$ and 
90$^\circ$ for the dipole-quadrupole cascades, whereas it decreases for 
the quadrupole-quadrupole or dipole-dipole ones. 
The theoretical values of several coincidence rates for the Euroball geometry have been already given 
in previous papers~\cite{as06,po11}. The method has been checked 
by correctly reproducing the expected angular correlations of $\gamma$-transitions 
having well known multipole orders and belonging to various 
fission fragments.

\section{Experimental results}

\subsection{Study of the odd-$Z$ isotones}

\subsubsection{Level scheme of $^{137}_{~55}$Cs\label{Cs137}}
The yrast states of $^{137}$Cs were obtained up to $\sim~5.5$~MeV from the 
analysis of the $^{248}$Cm spontaneous-fission study~\cite{br99}. More recently a
few transitions were added to this level scheme, using a $^{252}$Cf 
source~\cite{li07}. Nevertheless the spin-parity values were only based on
results of shell-model calculations.

\begin{figure*}[!htb]
\includegraphics[angle=90,width=17cm]{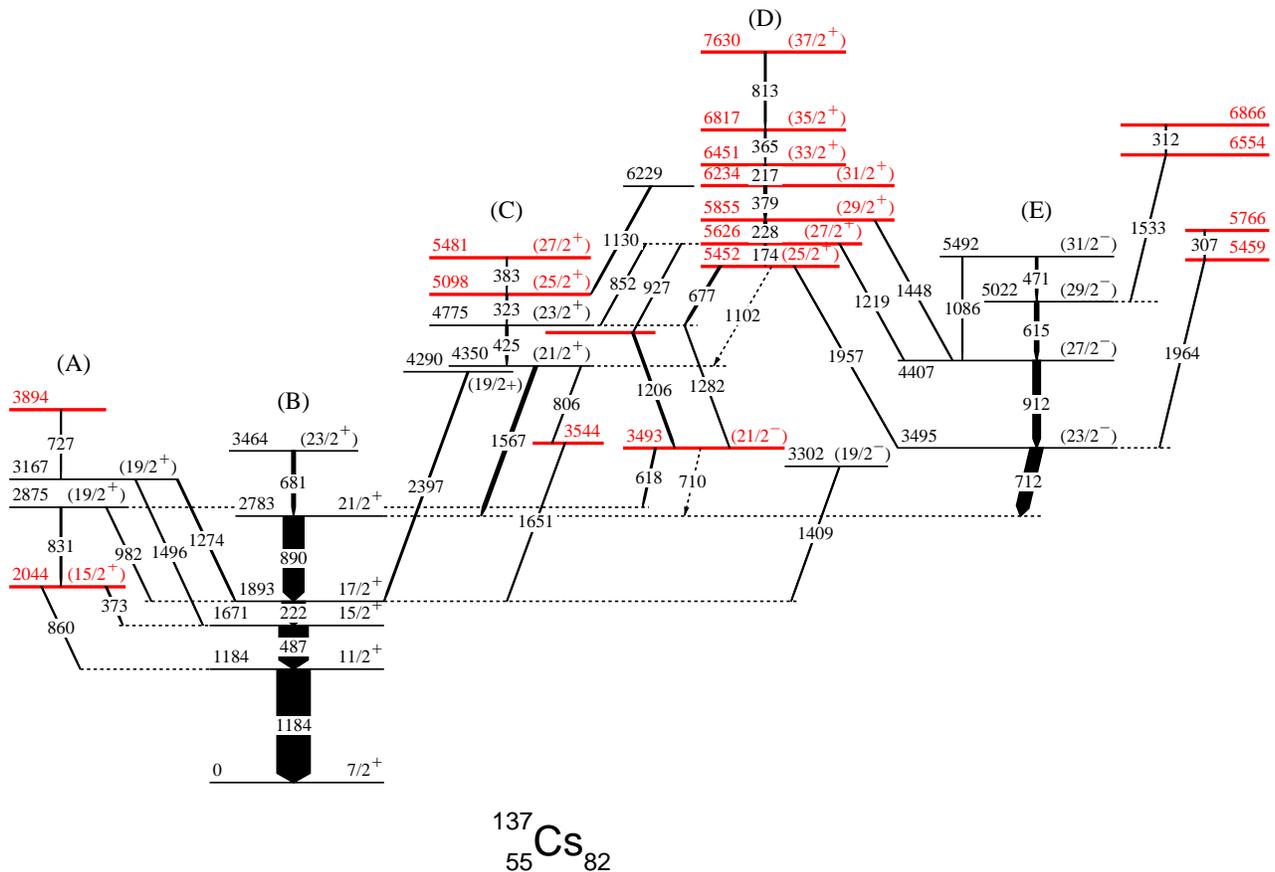}
\caption[]{(Color online) Level scheme of $^{137}$Cs established in this work. 
The colored states are new. The width of the arrows is proportional to the $\gamma$-ray intensity.
}
\label{schema_cs137}
\end{figure*}

In our work, $\gamma$ rays emitted by $^{137}$Cs are seen in the two  
fusion-fission reactions, with a slightly higher production in the (C+U) 
reaction compared to the (O+Pb) one, especially for the high-spin states.
The high-spin level scheme of $^{137}$Cs has been extended up to $\sim 7.6$~MeV 
excitation energy and spin ($37/2^+$), thanks to the observation of about 25 new 
transitions (see Fig.~\ref{schema_cs137}).

A 681-keV transition, already identified in the two previous experiments, is also
oberved in our work, being located just above the 2783-keV level (Structure B). The coincidence 
relationships of this $\gamma$ line involve the four yrast transitions, but we
could not observe any transition located above the 3464-keV level, 
although the  intensity of the 681-keV transition would have allowed us to see a
higher-spin part of the cascade. This level is likely populated by several high-energy
transitions which have escaped from our detection apparatus.

We do not confirm two of the six new transitions above the 1893-keV level, which
were proposed by the authors of Ref.~\cite{li07} in addition to the previous level
scheme~\cite{br99}. The 2223.6- and 1609.4-keV $\gamma$ rays are not observed in
coincidence with the first yrast transitions. Moreover the decay of the 3495-keV 
level toward the 3302-keV one by means of a 193-keV transition is not observed. 

\begin{figure}[!h]
\includegraphics*[width=8.5cm]{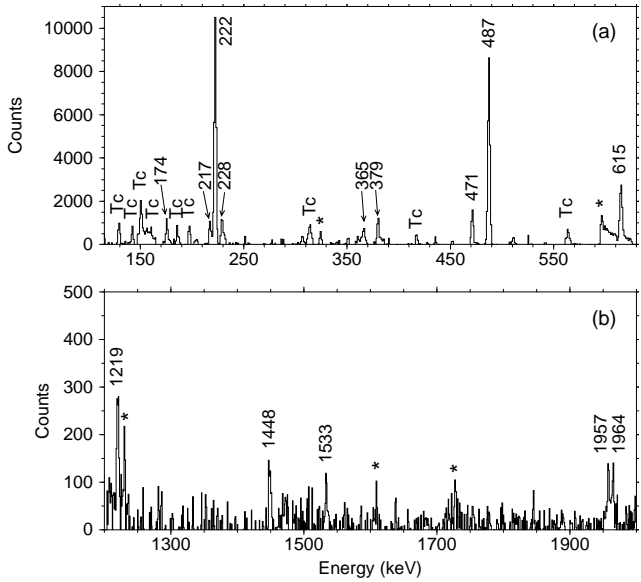}
\caption[]{Spectrum of $\gamma$ rays in coincidence with two transitions of $^{137}$Cs
showing the first transitions of Structure D in the low-energy part (a) and the links
between Structures D and E in the high-energy part (b). The first gate is set on the
712-keV transition and the second gate is set on one of the four yrast transitions (1184,
487, 222, and 890 keV). Transitions emitted by $^{103,104,105}$Tc, the complementary 
fragments of $^{137}$Cs, are labeled by Tc. The peaks marked with a star are contaminants.
}
\label{spectre_cs137}
\end{figure}

The most noticeable information obtained in this work is the observation of 
two structures composed of low-energy transitions (therefore presumably M1) located
above the 4350- and 5452-keV levels (Structures C and D in 
Fig.~\ref{schema_cs137}). The decays of their low-lying
states are very fragmented and involve high-energy transitions. An example of double-gated
spectrum showing the first transitions of Structure D and its links to Structure E is
given in Fig.~\ref{spectre_cs137}.

Angular correlations of successive $\gamma$ rays have been extracted for 
the most intense transitions of $^{137}$Cs. The experimental results are 
given in Table~\ref{correl_Cs}. The three first results indicate that the
transitions at 1184~keV, 487~keV, and 890~keV have the same multipole order, while
the last one indicates that the multipole order of the 222-keV transition is
different from that of the three others. An $E2$ character is ruled out for the  
222-keV transition, because the 1893-keV level would then be an isomeric state,
with a half-life of a few tens to a few hundreds of ns, at variance with the fact 
that the 1184-487-222~keV cascade is
not observed in the SAPhIR experiment. Thus the 222-keV transition is
assigned to be a dipole transition, while the 1184-, 487-, and 890-keV transitions
have an $E2$ character.  These arguments result in the spin values of
11/2$^+$, 15/2$^+$, 17/2$^+$, and 21/2$^+$ for the levels at 1184, 1671, 1893, and
2783~keV, belonging to Structure B.
\begin{table}[!h]
\caption{Coincidence rates between the low-lying $\gamma$ rays of $^{137}$Cs as 
a function of their relative angle of detection, normalized by the ones obtained 
around 75$^\circ$.}
\label{correl_Cs}
\begin{ruledtabular}
\begin{tabular}{cccc}
$\gamma$-$\gamma$ coincidence & R(22$^\circ$)$^{(a)}$ & R(46$^\circ$)$^{(a)}$ & R(75$^\circ$)$^{(a)}$\\
\hline
1184 - ~487  &  1.19(9)& 1.07(8)	&1.00(6)	\\
~890 - ~487  &  1.20(8)& 1.08(7)	&1.00(4)	\\
~890 - 1184  &  1.09(9)& 1.07(7)	&1.00(6)	\\
~222 - 1184  &  0.93(7)& 0.99(6)	&1.00(5)	\\
\end{tabular}
\end{ruledtabular}
\footnotetext[1]{The number in parentheses is the error in the 
last digit.}
\end{table}

The spin assignments of the higher-lying states are based on the following assumptions: 
(i) In the yrast decays, spin values increase with excitation energy, (ii)
The low-energy transitions have a $M1$ character. In addition the negative
parity proposed in Ref.~\cite{br99} to the states of the group labeled E because of the 
comparison with
calculations is in agreement with its single
link to the 21/2$^+$ state. The parity of the states of Structures C
and D will be discussed in Sect.~\ref{SM_odd}, from the comparison with theoretical
predictions.

The properties of the $^{137}$Cs transitions are given in Table~\ref{gammas_Cs137}.
The intensity values have been extracted 
as follows: First the ratio, I$_\gamma$(487~keV)/I$_\gamma$(1184~keV)$=0.88$,  
has been measured in a spectrum built with double gates set on its main 
complementary fragments ($^{103,105}$Tc) in the (C+U) experiment. Then, the 
spectrum gated simultaneously on the 1184-keV $\gamma$-ray and Tc transitions 
gives the 222-keV intensity relative to the 487-keV one. Finally all other 
intensities have been measured from spectra built with double gates set on 
$^{137}$Cs transitions.
\begin{table}[!h]
\caption{Properties of the transitions assigned to $^{137}$Cs observed in this work.}
\label{gammas_Cs137}
\begin{ruledtabular}
\begin{tabular}{rcccc}
$E_\gamma$\footnotemark[1](keV)  &  $I_\gamma$\footnotemark[1]$^,$\footnotemark[2]  &  $I_i^\pi \rightarrow I_f^\pi$  &$E_i$&$E_f$  \\
\hline
174.4(3)  &  4.5(22)   &   (27/2$^+$)  $\rightarrow$ (25/2$^+$)  &  5626.5  &  5452.2    \\
216.9(3)  &  5.5(20)   &    (33/2$^+$) $\rightarrow$ (31/2$^+$)  &  6451.3  & 6234.4     \\
221.8(3)  &  70(10)    &   17/2$^+$    $\rightarrow$ 15/2$^+$    &  1893.0  &  1671.2    \\
228.5(3)  &  8.2(25)   &   (29/2$^+$)  $\rightarrow$ (27/2$^+$)  &  5855.0  &  5626.5    \\
306.7(4)  &  1.0(5)    &           				&  5766.1  &  5459.4    \\
311.9(4)  &  1.0(5)    &           				&  6866.1  &  6554.2    \\
323.5(4)  &  3.8(19)   &   (25/2$^+$)  $\rightarrow$ (23/2$^+$)  &  5098.2  &  4774.7    \\
365.4(6)  &  4.0(20)   &     (35/2$^+$)$\rightarrow$ (33/2$^+$)  &  6816.7  & 6451.3     \\
372.6(4)  &  3.5(17)   &   (15/2$^+$)  $\rightarrow$ 15/2$^+$    &  2043.8  &  1671.2    \\
379.4(4)  &  8.4(25)   &   (31/2$^+$)  $\rightarrow$ (29/2$^+$)  &  6234.4  &  5855.0    \\
383.3(4)  &  3.5(17)   &    (27/2$^+$)  $\rightarrow$ (25/2$^+$) &  5480.5  &  5098.2    \\
424.9(4)  &  5.9(24)   &   (23/2$^+$)  $\rightarrow$ (21/2$^+$)  &  4774.7  &  4349.8    \\
470.6(4)  &  4.5(18)   &   (31/2$^-$)  $\rightarrow$ (29/2$^-$)  &  5492.2  &  5021.6    \\
486.8(3)  &  88(13)    &   15/2$^+$    $\rightarrow$ 11/2$^+$    &  1671.2  &  1184.4    \\
614.8(4)  &  11(3)     &   (29/2$^-$)  $\rightarrow$ (27/2$^-$)  &  5021.6  &  4406.8    \\
618.0(4)  &  3.4(17)   &   (21/2$^-$)  $\rightarrow$ (19/2$^+$)  &  3493.1  &  2875.1    \\
677.5(5)  &  4.6(18)   &   (25/2$^+$)  $\rightarrow$ (23/2$^+$)  &  5452.2  &  4774.7    \\
681.3(4)  &  10(3)     &   (23/2$^+$)  $\rightarrow$  21/2$^+$   &  3464.3  &  2783.0    \\
710.3(5)  &  $<$ 1     &   (21/2$^-$)  $\rightarrow$ (21/2$^+$)  &  3493.1  &  2783.0    \\
712.1(4)  &  37(7)     &   (23/2$^-$)  $\rightarrow$  21/2$^+$   &  3495.1  &  2783.0    \\
726.7(4)  &  0.8(4)    &               $\rightarrow$ (19/2$^+$)  &  3893.9  &  3167.2    \\
805.7(6)  &  0.5(2)    &   (21/2$^+$)  $\rightarrow$  		  &  4349.8  &  3544.3    \\
813.0(3)  &  3.7(18)   &   (37/2$^+$)  $\rightarrow$ (35/2$^+$)  &  7629.7  &  6816.7    \\
831.2(4)  &  3.5(17)   &   (19/2$^+$)  $\rightarrow$ (15/2$^+$)  &  2875.1  &  2043.8    \\
851.7(5)  &  2.1(10)   &   (27/2$^+$)  $\rightarrow$ (23/2$^+$)  &  5626.5  &  4774.7    \\
859.7(6)  &  0.7(3)    &   (15/2$^+$)  $\rightarrow$  11/2$^+$   &  2043.8  &  1184.4    \\
890.0(4)  &  62(9)     &   21/2$^+$    $\rightarrow$ 17/2$^+$    &  2783.0  &  1893.0    \\
911.7(4)  &  21(5)     &   (27/2$^-$)  $\rightarrow$ (23/2$^-$)  &  4406.8  &  3495.1    \\
927.2(5)  &  1.6(8)    &   (27/2$^+$)  $\rightarrow$ 	         &  5626.5  &  4699.6    \\
982.2(4)  &  2.5(12)   &   (19/2$^+$)  $\rightarrow$ 17/2$^+$    &  2875.1  &  1893.0    \\
1085.6(4) &  1.5(7)    &   (31/2$^-$)  $\rightarrow$ (27/2$^-$)  &  5492.2  &  4406.8    \\
1102.5(5) &  $<$ 1     &   (25/2$^+$)  $\rightarrow$ (21/2$^+$)  &  5452.2   & 4349.8     \\
1130.5(4) &  1.8(9)    &               $\rightarrow$ (25/2$^+$)  &  6228.7  &  5098.2    \\
1184.4(4) & 100(15)    &   11/2$^+$    $\rightarrow$  7/2$^+$    &  1184.4  &     0.0    \\
1206.5(4) &  1.8(9)    &    	   $\rightarrow$ (21/2$^-$)      &  4699.6  &  3493.1    \\
1219.4(4) &  1.9(9)    &   (27/2$^+$)  $\rightarrow$ (27/2$^-$)  &  5626.5  &  4406.8    \\
1274.3(4) &  3.7(18)   &   (19/2$^+$)  $\rightarrow$ 17/2$^+$    &  3167.2  &  1893.0    \\
1281.9(4) &  2.1(10)   &   (23/2$^+$)  $\rightarrow$ (21/2$^-$)  &  4774.7  &  3493.1    \\
1409.3(4) &  2.5(12)   &    (19/2$^-$)   $\rightarrow$ 17/2$^+$  &  3302.3  &  1893.0    \\
1448.5(5) &  1.6(8)    &   (29/2$^+$)  $\rightarrow$ (27/2$^-$)  &  5855.0  &  4406.8    \\
1495.7(6) &  1.3(6)    &   (19/2$^+$)  $\rightarrow$ 15/2$^+$    &  3167.2  &  1671.2    \\
1532.6(5) &  2.1(10)   &               $\rightarrow$ (29/2$^-$)  &  6554.2  &  5021.6    \\
1566.7(5) &  8.9(27)   &   (21/2$^+$)  $\rightarrow$  21/2$^+$   &  4349.8  &  2783.0    \\
1651.3(7) &  0.5(2)    &   	       $\rightarrow$ 17/2$^+$    &  3544.3  &  1893.0    \\
1957.2(7) &  1.4(7)    &   (25/2$^+$)  $\rightarrow$ (23/2$^-$)  &  5452.2  &  3495.1    \\
1964.3(8) &  1.7(9)    &               $\rightarrow$ (23/2$^-$)  &  5459.4  &  3495.1    \\
2397(1)   &  3.0(15)   &   (19/2$^+$)  $\rightarrow$ 17/2$^+$    &  4290    &  1893.0    \\
\end{tabular}
\end{ruledtabular}
\footnotetext[1]{The number in parentheses is the error in the last digit.}
\footnotetext[2]{The relative intensities are normalized to $I_\gamma(1184) = 100$.}
\end{table}

\subsubsection{Level scheme of $^{139}_{~57}$La\label{La139}}

\begin{figure*}[!htb]
\includegraphics[angle=90,width=14cm]{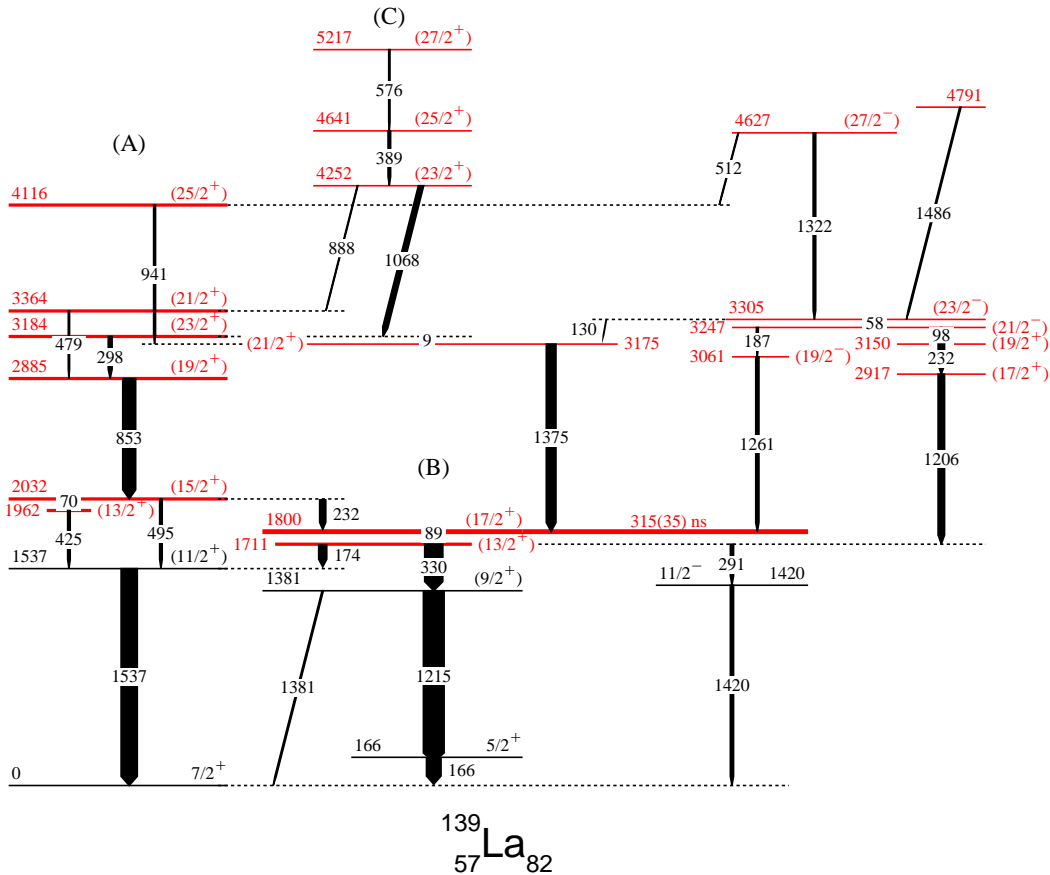}
\caption[]{(Color online) Level scheme of $^{139}$La established in this work.  
The colored states are new. The width of the arrows is indicative of the $\gamma$-ray intensity (see text).
}
\label{schema_la139}
\end{figure*}

High-spin states in $^{139}$La were scarcely known before our work. Four
states with $I$ = 9/2 and two states with $I$ = 11/2 have been proposed 
from $(n, n'\gamma)$ reaction, at an excitation energy below 1.6~MeV~\cite{nndc}, 
most of these states directly decaying to the ground state. Thus in order to 
identify the $\gamma$-ray cascades emitted by the high-spin states of $^{139}$La,
we have first used double-gated spectra on transitions of $^{101,102,103}$Nb 
isotopes~\cite{hw98}, the complementary fragments of $^{139}$La in the (C+U) 
fusion-fission experiment. 
We cannot confirm the existence of a $9/2^+$ state at 1219~keV, which should 
have been the yrast one, since we do not observe any of the 1219- or 1051-keV 
transitions~\cite{nndc}. However we confirm the $(9/2^+)$ state at 1381~keV, 
thanks to its decay toward the $5/2^+$ level (located at 166~keV), 
{\it via} the 1215-keV 
line which is in clear coincidence with the known 166-keV $5/2^+ \to 7/2^+$ 
transition. The 1215-keV transition is one of the 
strongest transition observed in coincidence when gating directly on the 
complementary fragments. 
The two other reported $9/2^+$ levels, being non-{\it yrast}, cannot be identified 
in our experiment.
Secondly, the 166-1215 two-gate spectrum easily reveals, in addition to Nb transitions, 
many new $\gamma$-lines which have to be located above the 1381-keV state of 
$^{139}$La. Finally,
all the coincidence relationships between these new transitions have been carefully
analyzed in order to build the level scheme shown in Fig.~\ref{schema_la139}. 

Two levels at very close excitation energies are reported in the literature at 
1537.7 and 1537.9~keV, with proposed spin assignments of ($11/2^+$) for the former,
which only decays to the ground state, and $7/2^+$ for the latter, which populates 
equally the ground state and the first excited state at 166~keV~\cite{nndc}. 
The 1537-keV $\gamma$-line is only present in our data in 
coincidence with the Nb transitions, meaning that only the ($11/2^+$) state has been
populated in our experiment. 

A second doublet of close-lying states is known at 1420.5~keV and 1420(12)~keV~\cite{nndc}. 
The first one, populated in the $\beta$-decay of $^{139}$Ba 
($I^\pi_{gs}=7/2^-$) with $log ft$=7.6, shows a main decay to the ground state and a weak
component to the first excited 
state (9\%). The second one was populated with a L=5 transfer in 
($^3$He,$d$), ($d$,$^3$He), and ($^7$Li, $^6$He) reactions~\cite{nndc}, thus assigned as
$I^\pi=11/2^-$ (from the $\pi h_{11/2}$ orbit). Its $\gamma$-decay, not measured, likely involves two
branches, the one toward the ground state with a $M2$ multipolarity and the one toward the
166-keV level with an $E3$ multipolarity, $i.e.$, similar to those measured in the 
neighboring isotones,  $^{141}$Pr and $^{143}$Pm~\cite{nndc}. Thus such a de-excitation is 
the same as that measured for the 1420.5-keV  level identified in the $\beta$-decay of 
$^{139}$Ba. This could indicate that there is only one state at 1420.5~keV. Nevertheless
the measured value of the direct $\beta$-feeding is too high for a second-forbidden non-unique
transition ($\Delta J=2$, $\Delta \pi=+$). Thus the results of $\beta$-decay and transfer 
reactions cannot be reconciled and two states very close in energy do exist in $^{139}$La. 
The 1420-keV state identified in the present work only decays toward the ground state, but
a weak decay to the 166-keV state cannot be excluded when taking into account the low
intensity measured for the 1420-keV line.  In Fig.~\ref{schema_la139}, we have chosen 
the value $I^\pi=11/2^-$, as discussed below.

\begin{figure}[h]
\includegraphics*[width=8.5cm]{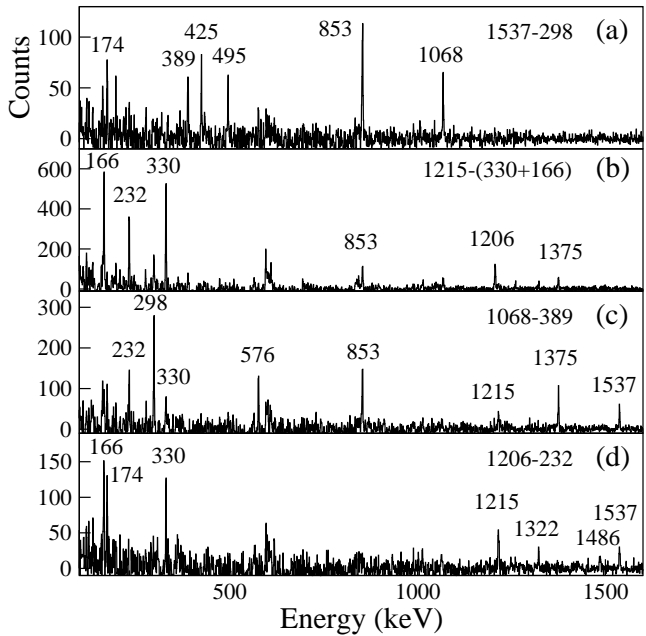}
\caption[]{Examples of double-gated spectra illustrating the most intense $^{139}$La transitions.
}
\label{spectres_La139}
\end{figure}

Noteworthy is the fact that the 232-keV transition has to be a doublet
in order to satisfy all its coincidence relationships, the first member is
located above the 1800-keV state and the second one above the 2917-keV state.
Some examples of double-gated spectra showing the most intense $^{139}$La 
transitions are displayed in Fig.~\ref{spectres_La139}.

\begin{figure}[h]
\includegraphics*[width=7cm]{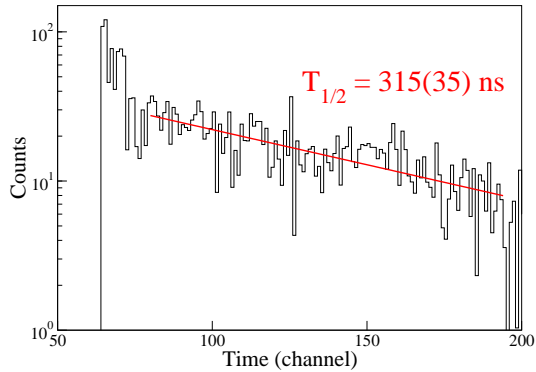}
\caption[]{(Color online) Time distribution between the detection of two fragments by SAPhIR 
and of a $\gamma$ transition by Euroball  (the $\gamma$ multiplicity 
being two) showing the half-life of the 1800-keV state of $^{139}$La: the  
$\gamma$ transition is either the 1215- or the 330-keV one, with a gate set
on the second $\gamma$, the 166-keV or the 330-/1215-keV transition.  
}
\label{TAC139La}
\end{figure}

By using the data from the SAPhIR experiment, the three cascades involved in the
decay of the 1800-keV level have been found to be delayed. The time distribution
between the detection of two fragments by SAPhIR and of the 1215-
or the 330-keV transition by Euroball is shown in Fig.~\ref{TAC139La}. In order to reduce
the background, we have selected the events containing an additional
$\gamma$-ray belonging to the 330-1215-166 cascade. Thus the half-life of the 1800-keV level
has been obtained, $T_{1/2}$=315(35)ns.

The level scheme of $^{139}$La includes five transitions with an energy lower 
than 100~keV. We have observed all their $\gamma$ lines but the 9- and the 58-keV 
transitions. The former is located between the levels at 3184~keV and 3175~keV because 
of some coincidence relationships. Indeed the spectrum
gated by the 1068- and 389-keV lines exhibits both the 1375-keV transition and the 298/853-keV
cascade [see Fig.~\ref{spectres_La139}(c)], while the spectrum
gated by 1322- and 130-keV lines only exhibits the 1375-keV transition. For the 58-keV 
transition, it is needed to account for all the coincidence relationships of the 1322-
and 1486-keV transitions. 
 
The determination of the transition intensities is rather delicate because of 
the isomeric state at 1800~keV which induces a loss in intensity  
$\sim 50\%$ for many transitions in the gated spectra, while some others 
keep their relative intensity, as they belong to decay paths by-passing the 
isomeric state. Therefore we do not give any intensity value in 
Table~\ref{gammas_La139}. On the other hand, the de-excitation branching ratios of  
levels, if any, can be determined as they do not suffer 
the drawback mentioned above, they are given in Table~\ref{branching_La139}.
\begin{table}[!h]
\caption{Properties of the transitions assigned to $^{139}$La observed in this work. 
}
\label{gammas_La139}
\begin{ruledtabular}
\begin{tabular}{rccc}
$E_\gamma$\footnotemark[1](keV)&   $I_i^\pi \rightarrow I_f^\pi$  &$E_i$&$E_f$  \\
\hline
9.0\footnotemark[2]&    (23/2$^+$)  $\rightarrow$  (21/2$^+$)&  3183.7   & 3174.7   \\
57.7\footnotemark[2]&    (23/2$^-$)  $\rightarrow$  (21/2$^-$) &  3305.0  & 3247.3   \\
70.0(3)         &    (15/2$^+$)  $\rightarrow$  (13/2$^+$) &  2032.1  &  1962.2    \\
88.7(5)         &    (17/2$^+$)  $\rightarrow$  (13/2$^+$) &  1799.8  &  1711.1   \\
97.6(5)	        &    (21/2$^-$)  $\rightarrow$  (19/2$^+$) &  3247.3   & 3149.7   \\
130.3(3)        &    (23/2$^-$)  $\rightarrow$  (21/2$^+$) &  3305.0  &  3174.7   \\
165.6(3)       &    5/2$^+$     $\rightarrow$  7/2$^+$     &  165.6   &     0.0   \\
174.0(3)       &    (13/2$^+$)  $\rightarrow$  (11/2$^+$)  &  1711.1  &  1537.0  \\
186.7(4)       &    (21/2$^-$)  $\rightarrow$  (19/2$^-$)  &  3247.3  &  3060.6  \\
232.3(4)       &    (15/2$^+$)  $\rightarrow$  (17/2$^+$)  &  2032.1  &  1799.8   \\
232.3(4)       &    (19/2$^+$)  $\rightarrow$  (17/2$^+$)  &  3149.7  &  2917.4   \\
291.3(3)       &    (13/2$^+$)  $\rightarrow$  11/2$^-$    &  1711.1  &  1420.0  \\
298.4(3)       &    (23/2$^+$)  $\rightarrow$  (19/2$^+$)  &  3183.7  &  2885.3   \\
330.3(3)       &    (13/2$^+$)  $\rightarrow$  (9/2$^+$)   &  1711.1  &  1380.8  \\
388.9(3)       &    (25/2$^+$)  $\rightarrow$  (23/2$^+$)  &  4640.6  &  4251.7  \\
425.2(3)       &    (13/2$^+$)  $\rightarrow$  (11/2$^+$)  &  1962.2  &  1537.0  \\
478.6(4)       &    (21/2$^+$)  $\rightarrow$  (19/2$^+$)  &  3363.9  &  2885.3   \\
495.0(3)       &    (15/2$^+$)  $\rightarrow$  (11/2$^+$)  &  2032.1  &  1537.0   \\
512(1)         &    (27/2$^-$)  $\rightarrow$  (25/2$^+$)  &  4627.3  &  4115.6  \\
576.4(4)       &    (27/2$^+$)  $\rightarrow$  (25/2$^+$)  &  5217.0  &  4640.6   \\
853.2(4)       &    (19/2$^+$)  $\rightarrow$  (15/2$^+$)  &  2885.3  &  2032.1  \\
887.8(4)       &    (23/2$^+$)  $\rightarrow$  (21/2$^+$)  &  4251.7  &  3363.9 \\
940.9(4)       &    (25/2$^+$)  $\rightarrow$  (21/2$^+$)  &  4115.6  &  3174.7  \\
1068.0(4)      &    (23/2$^+$)  $\rightarrow$  (23/2$^+$)  &  4251.7  &  3183.7 \\
1215.2(5)      &    (9/2$^+$)   $\rightarrow$  5/2$^+$     &  1380.8  &   165.6  \\
1206.3(5)      &    (17/2$^+$)  $\rightarrow$  (13/2$^+$)  &  2917.4  &  1711.1  \\
1260.8(5)      &    (19/2$^-$)  $\rightarrow$  (17/2$^+$)  &  3060.6  &  1799.8  \\
1322.3(5)      &    (27/2$^-$)  $\rightarrow$  (23/2$^-$)  &  4627.3  &  3305.0   \\
1374.9(4)      &    (21/2$^+$)  $\rightarrow$  (17/2$^+$)  &  3174.7  &  1799.8   \\
1380.9(6)      &    (9/2$^+$)   $\rightarrow$  7/2$^+$     &  1380.8  &    0.0   \\
1420.0(5)      &    11/2$^-$    $\rightarrow$  7/2$^+$     &  1420.0  &    0.0  \\
1485.6(6)      &    		$\rightarrow$  (23/2$^-$)  &  4790.6  &  3305.0  \\
1537.0(4)    &    (11/2$^+$)  $\rightarrow$  7/2$^+$       &  1537.0  &  0.0    \\

\end{tabular}
\end{ruledtabular}
\footnotetext[1]{The number in parentheses is the error in the last digit.}
\footnotetext[2]{Not observed, inferred from the coincidence relationships.}

\end{table}
\begin{table}[!h]
\caption{Branching ratios of the states of $^{139}$La measured in this work. 
}
\label{branching_La139}
\begin{ruledtabular}
\begin{tabular}{crrr}

State  &  $E$~-~{\it $I_{tot}$}\footnotemark[1]$^,$\footnotemark[2]  	
&$E$~-~{\it $I_{tot}$}\footnotemark[1]$^,$\footnotemark[2]  
&$E$~-~{\it $I_{tot}$\footnotemark[1]$^,$\footnotemark[2]}\\
 (keV) &  (keV)-(\%)  				&(keV)-(\%)&(keV)-(\%)\\
\hline
1381		&1215~-~{\it 95(9})& 1381~-~{\it ~5(2)} &\\

1711   	& 174~-~{\it 35(6)} &  291~-~{\it 10(3)} & 330~-~{\it 55(8)}\\

2032   	& 70~-~{\it 20(4)} & 232~-~{\it 60(8)} &495~-~{\it 20(4)}\\

3184  	& 298~-~{\it 55(8)} & 9~-~{\it 45(7)} &\\

3247		& 98~-~{\it 55(8)} & 187~-~{\it 45(7)} &\\

3305		&58~-~{\it 70(9)} & 130~-~{\it 30(5)} &\\

4252		&888~-~{\it 14(3)} & 1068~-~{\it 86(9)} &\\

4627		&512~-~{\it 10(3)} & 1322~-~{\it 90(9)} &\\
		
\end{tabular}
\end{ruledtabular}
\footnotetext[1]{The intensity balances used to measure the branching ratios 
take into account the conversion electrons of the low-energy transitions.}
\footnotetext[2]{The number in parentheses is the error in the last digit.}
\end{table}

Angular correlations of transitions emitted by $^{139}$La could not be measured due 
to their too weak intensities in the (O+Pb) experiment. Spin and parity assignments 
have been therefore deduced by using the following arguments: (i) The spin values 
increase with excitation energy, (ii) The high-energy (low-energy) transitions likely have an 
$E2~(M1)$ character, (iii) The measured branching ratios as well as the existence 
or the absence of cross-over transitions place some conditions on the 
multipolarities. Starting from the two values already known, $I^\pi$=7/2$^+$ for 
the ground state and $I^\pi$=5/2$^+$ for the first excited state~\cite{nndc}, we 
have obtained the values given in parentheses in Fig.~\ref{schema_la139}, some of 
them are discussed now. 

First of all, it is important to note that the level scheme of $^{139}$La  does 
not resemble the one of $^{137}$Cs, particularly its 5/2$^+$ state at 166~keV
takes part to the yrast structure, while the 5/2$^+$ state of $^{137}$Cs, 
being higher in energy (455~keV), is not observed in our work (see Fig.~\ref{schema_cs137}). 
On the other hand, the low-lying
part of the level scheme of $^{139}$La can be compared to the one of $^{141}$Pr, 
which exhibits both the 5/2$^+$ state and the 7/2$^+$ one in the in-beam 
studies~\cite{pr81}. This leads the spin values of the 1381-, 1537- and 1711-keV
states to be 9/2$^+$, 11/2$^+$, and 13/2$^+$, respectively. The 13/2$^+$ state of 
$^{141}$Pr also decays to the 11/2$^-$ level at 1118~keV. Thus we 
propose that the 1420-keV state measured in the present work is the 11/2$^-$ 
level already identified in transfer reactions, as the spin value
assigned to the other level, observed in the $\beta$-decay, $I^\pi$=5/2$^+$ or 
7/2$^+$~\cite{nndc} is too low. For the isomeric state at 1800~keV, we choose 
$I^\pi$=17/2$^+$, assuming that the 89-keV transition is $E2$. This leads to 
$B(E2; 17/2^+\rightarrow13/2^+)=81(9)~e^2fm^4$, $i.e.$, 1.9(2)~$W.u.$.
The yrast cascades of $^{141}$Pr and $^{143}$Pm also contain one delayed $E2$ 
transition which exhibits a similar value of hindrance, 0.87(10) $W.u.$ and 
1.52(6) $W.u.$, respectively. Nevertheless the spin value of their isomeric 
states is one unit less, 15/2$^+$. This change is due to the number of protons 
occupying the two orbits close to the Fermi level, namely $\pi g_{7/2}$ and $\pi d_{5/2}$.
For the decay of the 3184-keV state of $^{139}$La, the fact that a 9-keV transition 
competes with a 298-keV one
implies that the former is $M1$ while the latter has to be hindered, such as
expected for an $E2$ transition of low energy. Similar argument is used for the choice of  
the parity value of the 3305-keV level: The 58-keV transition is assumed to be $M1$ while 
the 130-keV transition would be $E1$, as imposed by their relative branching ratios
(Table~\ref{branching_La139}).  
 
\subsection{Study of the even-$Z$ isotones}
\subsubsection{Level scheme of $^{136}_{~54}$Xe\label{Xe136}}
\begin{figure}[!htb]
\includegraphics[width=8.6cm]{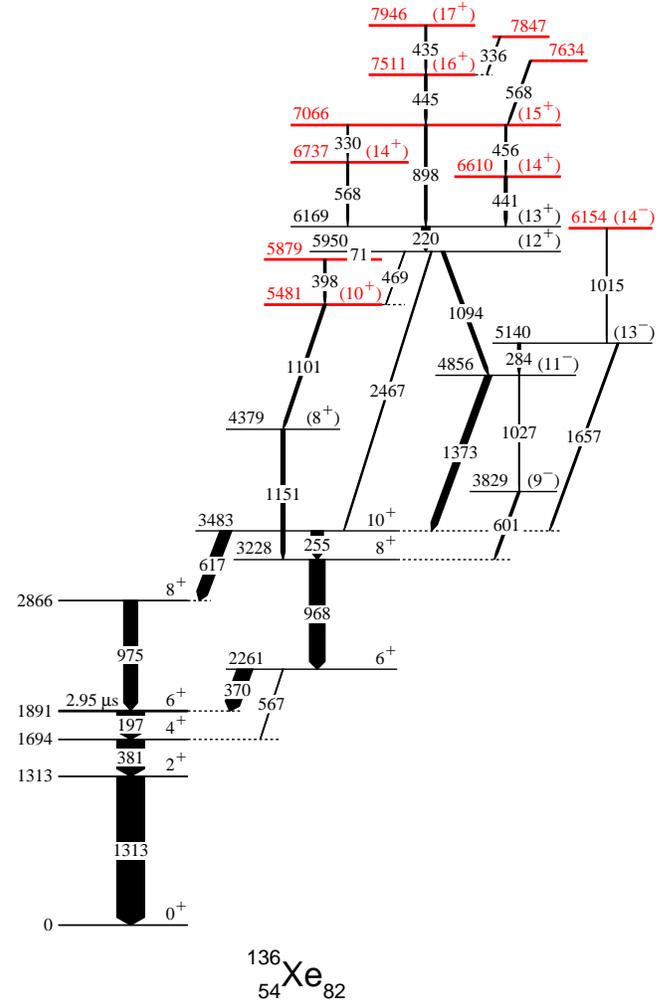}
\caption[]{(Color online) Level scheme of $^{136}$Xe established in this work.  
The colored states are new. The width of the arrows is proportional to the $\gamma$-ray intensity.
The half-life value of the 6$^+_1$ state is from Ref.~\cite{nndc}.
}
\label{schema_xe136}
\end{figure}

The yrast excitations of $^{136}$Xe, studied using a $^{248}$Cm source, were
identified up to an excitation energy of 6.2~MeV and a spin value of 
(13$^+$)~\cite{da99}. Since $^{136}$Xe is produced in both reactions used in
the present work and the fusion-fission process brings higher angular momenta 
than that obtained in 
the spontaneous fission of actinides, we have looked for new $\gamma$-ray 
cascades emitted by $^{136}$Xe. The level scheme built from these analyses is shown
in Fig.~\ref{schema_xe136}.
We have gathered in Table~\ref{gammas_Xe136} the properties of all the transitions
assigned to $^{136}$Xe from this work.
\begin{table}[!h]
\caption{Properties of the transitions assigned to $^{136}$Xe observed in this 
experiment.}\label{gammas_Xe136}
\begin{ruledtabular}
\begin{tabular}{rcccc}
$E_\gamma$\footnotemark[1](keV)& $I_\gamma$\footnotemark[1]$^,$\footnotemark[2]&  $I_i^\pi \rightarrow I_f^\pi$  &$E_i$&$E_f$  \\
\hline
70.7\footnotemark[3]&  6.2(19)\footnotemark[4] &(12$^+$)$\rightarrow$(11$^+$)&5949.5 &5878.8 \\
196.8(3)  & -      &    6$^+$   $\rightarrow$  4$^+$       &1890.8 	&1694.0	\\
219.5(3)  & 29(6)   &    (13$^+$)  $\rightarrow$  (12$^+$)  &6169.0 	&5949.5	\\
254.6(3)  & 44(7)   &     10$^+$   $\rightarrow$   8$^+$    &3482.8	&3228.1	\\
284.0(4)  & 8.8(26)  &    (13$^-$)  $\rightarrow$  (11$^-$)  &5139.8	&4855.8	\\
329.8(4)  & 7(2)  &    (15$^+$)  $\rightarrow$  (14$^+$)  &7066.3	&6736.5	\\
336.4(4)  & 3.4(14)  &            $\rightarrow$  (16$^+$)    &7847.2	&7510.8	\\
369.7(3)  & 53(8)   &    6$^+$   $\rightarrow$  6$^+$       &2260.5	&1890.8	\\
381.2(2)  & -      &    4$^+$   $\rightarrow$  2$^+$       &1694.0	&1312.8	\\
398.2(4)  & 7.8(23)  &  (11$^+$)  $\rightarrow$  (10$^+$) &5878.8	&5480.6	\\
435.4(4)  & 5.1(15)  &    (17$^+$)  $\rightarrow$  (16$^+$)  &7946.2	&7510.8	\\
441.2(3)  & 10(3)   &    (14$^+$)  $\rightarrow$  (13$^+$)  &6610.3	&6169.0	\\
444.5(4)  & 8.5(25)  &    (16$^+$)  $\rightarrow$  (15$^+$)  &7510.8	&7066.3	\\
455.9(4)  & 7(2)  &    (15$^+$)  $\rightarrow$  (14$^+$)  &7066.3	&6610.3	\\
469.1(5)  & 1.6(8)  &    (12$^+$)  $\rightarrow$ (10$^+$)    &5949.5	&5480.6	\\
567.0(5)  & 3(2)    &    6$^+$  $\rightarrow$  4$^+$        &2260.5	&1694.0	\\
567.5(5)  & 7(2)    &    (14$^+$)  $\rightarrow$  (13$^+$)  &6736.5	&6169.0	\\
568.0(5)  & 7(2)    &      $\rightarrow$  (15$^+$)          &7634.3	&7066.3	\\
600.8(4)  & 8(3)    &   (9$^-$)  $\rightarrow$   8$^+$      & 3828.9    &3228.1	\\
617.0(3)  & 39(6)   &     10$^+$   $\rightarrow$  (8$^+$)   &3482.8	&2865.9	\\
897.5(4)  & 7(2)  &    (15$^+$)  $\rightarrow$  (13$^+$)  &7066.3	&6169.0	\\
967.6(3)  & 56(8)   &     8$^+$   $\rightarrow$  6$^+$      &3228.1	&2260.5	\\
975.1(3)  & 47(7)   &    (8$^+$)  $\rightarrow$  6$^+$      &2865.9	&1890.8	\\
1014.6(4)  & 6(3)      &    (14$^-$)  $\rightarrow$  (13$^-$)&6154.4	&5139.8	\\
1027.1(4)  & 4(2)     &   (11$^-$)  $\rightarrow$  (9$^-$)    &4855.8   &3828.9\\
1093.7(3)  & 11(3)  &    (12$^+$)  $\rightarrow$  (11$^-$)  &5949.5	&4855.8	\\
1101.3(3)  & 9.7(29) &   (10$^+$)  $\rightarrow$  (8$^+$)   &5480.6	&4379.3	\\
1151.2(3)  & 13(3)  &    (8$^+$)  $\rightarrow$   8$^+$    &4379.3	&3228.1	\\
1312.8(2)  & -     &    2$^+$  $\rightarrow$  0$^+$        &1312.8	&0.0	\\
1373.0(4)  & 23(5)  &    (11$^-$)  $\rightarrow$   10$^+$   &4855.8	&3482.8	\\
1657.0(5)  & 6(3) &    (13$^-$)  $\rightarrow$     10$^+$  &5139.8	&3482.8	\\
2467.2(5)  & 5.0(25) &    (12$^+$)  $\rightarrow$  10$^+$   &5949.5	&3482.8	\\
\end{tabular}
\end{ruledtabular}
\footnotetext[1]{The number in parentheses is the error in the last digit.}
\footnotetext[2]{The relative intensities are normalized to the sum of the populations of 
the 6$^+_1$ isomeric
state, $I_\gamma(370)+I_\gamma(975) = 100$.}
\footnotetext[3]{Not observed, inferred from the coincidence relationships.}
\footnotetext[4]{This number is I$_{tot}$ and not I$_\gamma$.}
\end{table}

The level scheme has been extended up to 7.9~MeV excitation energy by
means of several $\gamma$ rays in cascade. Moreover three new decay paths of the 
(12$^+$) state at 5950~keV have been observed. The spin values of states lying
below 2.3~MeV excitation energy are those proposed in the previous 
works~\cite{nndc}. In addition we have measured the angular correlations of 
a few $\gamma$ rays of $^{136}$Xe lying above its long-lived isomeric state (see
Table~\ref{correl_Xe}). 
\begin{table}[!h]
\caption{Coincidence rates between a few $\gamma$ rays of $^{136}$Xe lying above
its long-lived isomeric state, as 
a function of their relative angle of detection, normalized by the ones obtained 
around 75$^\circ$.}
\label{correl_Xe}
\begin{ruledtabular}
\begin{tabular}{cccc}
$\gamma$-$\gamma$ coincidence & R(22$^\circ$)$^{(a)}$ & R(46$^\circ$)$^{(a)}$ & R(75$^\circ$)$^{(a)}$\\
\hline
968 - 370  &  1.5(3)  & 1.2(1)  & 1.00(6)	\\
255 - 968  &  1.2(1)  & 1.0(1)	& 1.00(5)	\\
\end{tabular}
\end{ruledtabular}
\footnotetext[1]{The number in parentheses is the error in the 
last digit.}
\end{table}
The results indicate that the 968- and 255-keV transitions have an $E2$
character, knowing that the 370-keV $\gamma$-ray is a dipole transition linking
two states having the same spin value~\cite{nndc}. It has to be noticed that a 
new $\gamma$-ray linking
the 2261-keV level to the 4$^+$ state at 1694~keV has been observed, which is in
good agreement with the involved spin values. Finally, the 2866-keV state has 
$I^\pi=8^+$, as it is directly populated by a 10$^+$ state and it decays 
toward a 6$^+$ state (see Fig.~\ref{schema_xe136}).
For the highest-spin part, we have assumed that the spin values increase with 
excitation energy and that the low-energy transitions have a $M1$ character.

We have looked for isomeric states in $^{136}$Xe by using the data registered with the
SAPhIR detector. Only one $\gamma$-ray cascade has been found to be delayed, that
decaying the known isomeric state at 1891~keV.

\subsubsection{Level scheme of $^{138}_{~56}$Ba\label{Ba138}}

\begin{figure*}[!ht]
\includegraphics[angle=90, width=15cm]{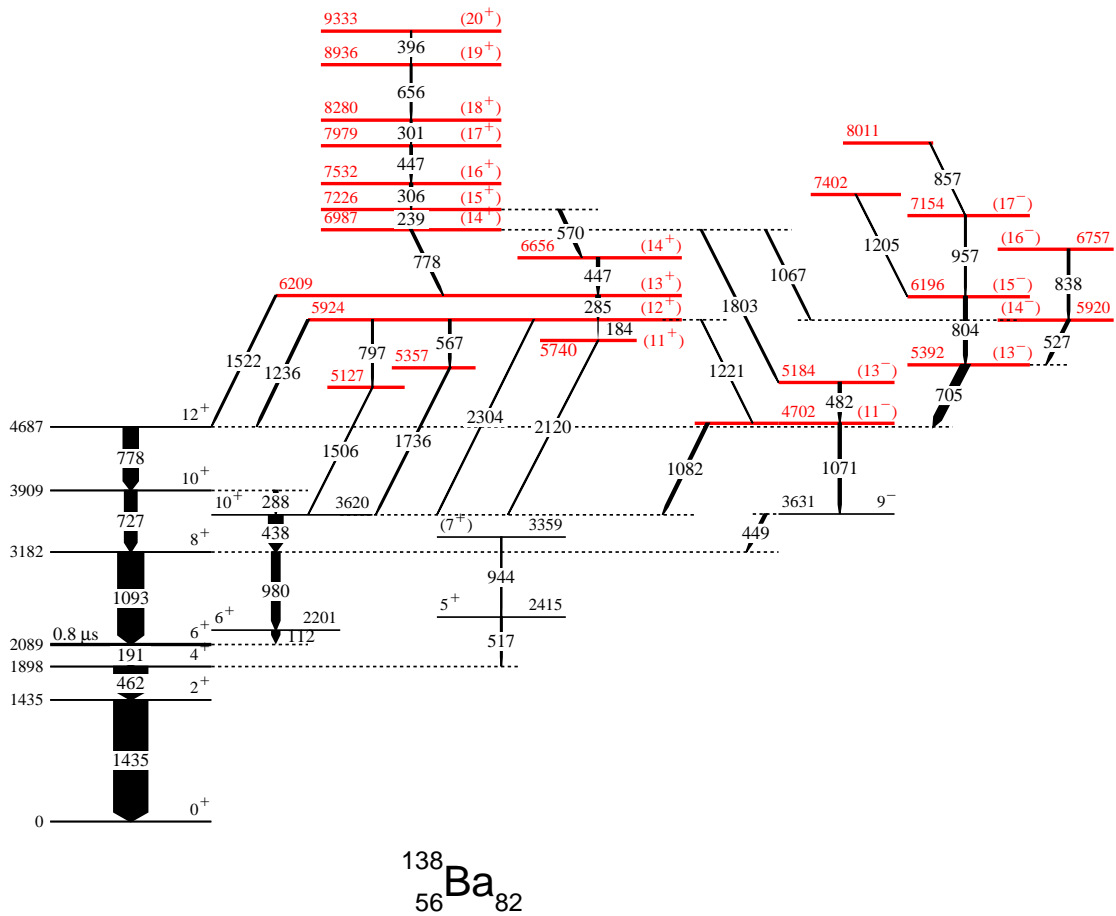}
\caption[]{(Color online) Level scheme of $^{138}$Ba established in this work. The
colored states are new. 
The width of the arrows is proportional to the $\gamma$-ray intensity.
The half-life value of the 6$^+$ state is from Ref.~\cite{nndc}.
}
\label{schema_ba138}
\end{figure*}

Previous information regarding the medium-spin states of $^{138}$Ba comes from results of the
($\alpha$,2n) reaction, where levels with spin values up to 12~$\hbar$ and excitation
energies of 4.7~MeV were identified~\cite{pr87}. All the yrast levels have been confirmed by the analyses of
both data sets of the present work. Moreover, the spectra doubly-gated on the known
transitions allowed us to identify many new $\gamma$-lines which extend the level scheme up to 
9.3~MeV excitation energy. 
We have gathered in Table~\ref{gammas_Ba138} the properties of all the transitions
assigned to $^{138}$Ba from this work and its level scheme is drawn in Fig.~\ref{schema_ba138}.
\begin{table}[!h]
\caption{Properties of the transitions assigned to $^{138}$Ba observed in this work.}
\label{gammas_Ba138}
\begin{ruledtabular}
\begin{tabular}{rcccc}
$E_\gamma$\footnotemark[1](keV)& $I_\gamma$\footnotemark[1]$^,$\footnotemark[2]&  $I_i^\pi \rightarrow I_f^\pi$  &$E_i$&$E_f$  \\
\hline
112.1(5)  &  27(5)    & 6$^+$    $\rightarrow$  6$^+$  	  &2201.4     &2089.3 	\\
183.7(5)  &  3.3(13)  & (12$^+$) $\rightarrow$  (11$^+$)  &5923.7     &5740.1 	\\
191.5(3)  &  -        & 6$^+$    $\rightarrow$   4$^+$ 	  &2089.3     &1897.8 	\\
239.0(4)  &  2.2(10)  & (15$^+$) $\rightarrow$  (14$^+$)  &7225.9     &6986.9 	\\
285.4(3)  &  12(4)    & (13$^+$) $\rightarrow$  (12$^+$)  &6209.0     &5923.7 	\\
288.2(3)  &  12(4)    & 10$^+$   $\rightarrow$  10$^+$    &3908.6     &3620.3 	\\
301.4(4)  &  4.4(18)  & (18$^+$) $\rightarrow$  (17$^+$)  &8280.1     &7978.7 	\\
306.1(3)  &  7.3(22)  & (16$^+$) $\rightarrow$  (15$^+$)  &7532.0     &7225.9 	\\
396.1(5)  &  2.1(10)  & (20$^+$) $\rightarrow$   (19$^+$) &9332.6     &8936.5 	\\
438.3(3)  &  40(6)    & 10$^+$   $\rightarrow$  8$^+$  	  &3620.3     &3182.0 	\\
446.7(3)  &  7.4(22)  & (14$^+$) $\rightarrow$ (13$^+$)   &6655.7     &6209.0 	\\
446.7(5)  &  5.6(17)  & (17$^+$) $\rightarrow$ (16$^+$)   &7978.7     &7532.0 	\\
449.1(3)  &  9.6(29)  &  9$^-$   $\rightarrow$ 8$^+$      &3631.1     &3182.0 	\\
462.4(3)  &  -        & 4$^+$    $\rightarrow$  2$^+$  	  &1897.8     &1435.4 	\\
481.8(3)  &  8.3(25)  & (13$^-$) $\rightarrow$  (11$^-$)  &5184.2     &4702.4 	\\
516.7(4)  &  3.2(13)  & 5$^+$    $\rightarrow$   4$^+$ 	  &2414.5     &1897.8 	\\
527.4(4)  &  6.9 (20) & (14$^-$) $\rightarrow$   (13$^-$)  &5919.6     &5392.2 	\\
567.3(3)  &  6.1(18)  & (12$^+$) $\rightarrow$    	  &5923.7     &5356.8 	\\
570.1(3)  &  7.1(21)  & (15$^+$) $\rightarrow$  (14$^+$)  &7225.9     &6655.7 	\\
656.4(5)  &  3.2(15)  & (19$^+$) $\rightarrow$ (18$^+$)   &8936.5     &8280.1 	\\
705.2(3)  &  23(5)    & (13$^-$) $\rightarrow$ 12$^+$     &5392.2     &4687.0 	\\
726.7(3)  &  35(7)    & 10$^+$   $\rightarrow$   8$^+$ 	  &3908.6     &3182.0 	\\
778.0(4)  &  5.4(19)  & (14$^+$) $\rightarrow$ (13$^+$)   &6986.9     &6209.0 	\\
778.4(3)  &  41(6)    & 12$^+$   $\rightarrow$ 10$^+$     &4687.0     &3908.6 	\\
797.1(4)  &  3.0(15)  & (12$^+$) $\rightarrow$    	  &5923.7     &5126.6 	\\
804.2(3)  &  9.4(28)  & (15$^-$) $\rightarrow$   (13$^-$) &6196.4     &5392.2 	\\
837.8(4)  &  5.1(18)  &	(16$^-$) $\rightarrow$   (14$^-$) &6757.4     &5919.6   \\
856.9(5)  &  2.4(12)  &  	 $\rightarrow$   (17$^-$)  &8010.7     &7153.8 	\\
944.0(5)  &  1.7(8)   & (7$^+$)  $\rightarrow$ 5$^+$   	  &3358.5     &2414.5 	\\
957.4(5)  &  3.1(15)  &  (17$^-$)$\rightarrow$ (15$^-$)    &7153.8     &6196.4 	\\
980.3(3)  &  23(5)    & 8$^+$    $\rightarrow$ 6$^+$   	  &3182.0     &2201.4 	\\
1067.1(5) &  3.5(15)  &	(14$^+$) $\rightarrow$ (14$^-$)   &6986.9     &5919.6  \\  
1071.3(3)  &  7.8(23) & (11$^-$) $\rightarrow$  9$^-$     &4702.4     &3631.1 	\\
1082.1(3)  &  7.8(23) & (11$^-$) $\rightarrow$  10$^+$    &4702.4     &3620.3 	\\
1092.7(3)  &  73(11)  & 8$^+$    $\rightarrow$  6$^+$  	  &3182.0     &2089.3 	\\
1205.2(5)  &  2(1)    &  	 $\rightarrow$  (15$^-$)   &7401.6     &6196.4 	\\
1221.2(5)  &  2.7(13) & (12$^+$) $\rightarrow$ (11$^-$)   &5923.7     &4702.4 	\\
1236.4(4)  &  6.0(18) & (12$^+$) $\rightarrow$ 12$^+$     &5923.7     &4687.0 	\\
1435.4(4)  &  -       & 2$^+$    $\rightarrow$ 0$^+$   	  &1435.4     & 0.0	\\
1506.3(5)  &  2.4(12) &       $\rightarrow$  10$^+$    &5126.6     &3620.3 	\\
1521.9(5)  &  4.2(15) & (13$^+$) $\rightarrow$  12$^+$    &6209.0     &4687.0 	\\
1736.4(5)  &  4.9(17) &       $\rightarrow$  10$^+$    &5356.8     &3620.3 	\\
1802.6(6)  &  2.9(14) & (14$^+$) $\rightarrow$   (13$^-$) &6986.9     &5184.2 	\\
2119.8(8)  &  2.1(10) & (11$^+$) $\rightarrow$ 10$^+$     &5740.1     &3620.3 	\\
2303.6(8)  &  1.4(7)  & (12$^+$) $\rightarrow$  10$^+$    &5923.7     &3620.3 	\\
\end{tabular}
\end{ruledtabular}
\footnotetext[1]{The number in parentheses is the error in the last digit.}
\footnotetext[2]{The relative intensities are normalized to the sum of the populations of 
the 6$^+_1$ isomeric
state, $I_\gamma(112)+I_\gamma(1093) = 100$.}
\end{table}

\begin{figure}[!h]
\includegraphics[width=8.5cm]{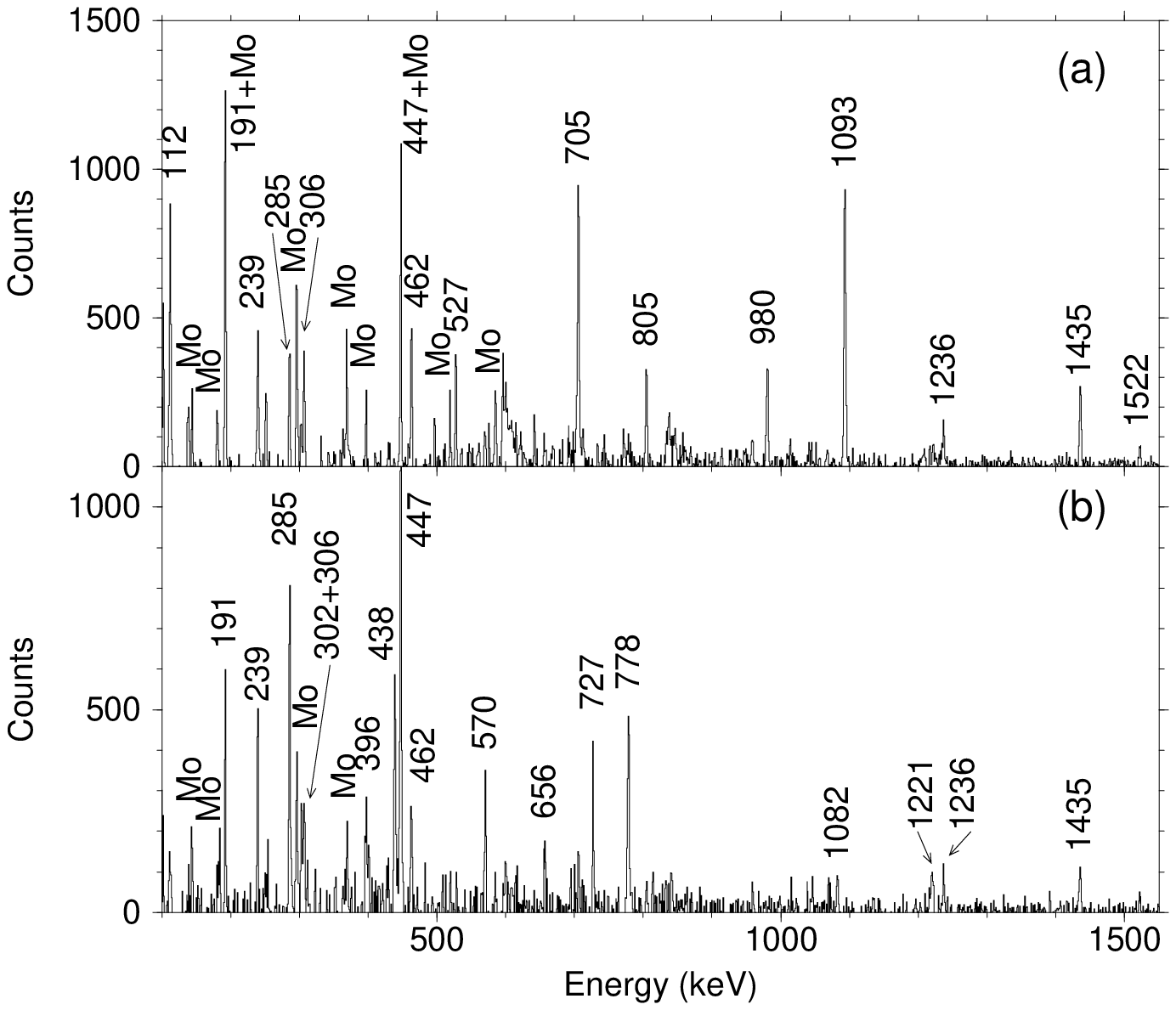}
\caption[]{Spectra of $\gamma$ rays  in coincidence with two transitions of 
$^{138}$Ba: (a) with the 727- and the 778-keV transitions; 
(b) with the 1093- and the 570- or 306-keV transitions. Transitions emitted by
$^{101-104}$Mo, the complementary fragments of $^{138}$Ba, are labeled by Mo.
}
\label{spectres_ba138}
\end{figure}

First, several transitions have been found to populate the 12$^+$ state at 4687~keV. The
most intense is the 705-keV transition [see the spectrum of 
Fig.~\ref{spectres_ba138}(a)] which defines the new level at 5392~keV. Above it, the
gamma intensity is spread over several lines. 
Secondly, a new structure has been built above the 9$^-$ state already known from 
the ($\alpha$,2n) reaction (see the right part of Fig.~\ref{schema_ba138}).
Thirdly, the high-energy part of the yrast structure of $^{138}$Ba consists of ten 
transitions in mutual coincidences which are located between 5740 keV and 9333 keV.
The three first states of this new band show many decay paths involving states
lying in the excitation-energy range [3620-5357]~keV, indicating a large change of 
configuration, which will be discussed in Sect.~\ref{discuss}.

For the spin assignments, we have first used the results of the previous 
experiment~\cite{pr87},
particularly the angular distributions and linear polarizations. It is worth
mentioning that the ($\alpha$,2n) reaction populates both yrast and yrare states of
the produced nuclei. Thus in many cases, the knowledge of the transition 
multipolarity was not enough to determine unambiguously the spin and parity values 
of the decaying state. For instance, even though the multipolarity of the 
449-keV transition was measured to be stretched $E1$ (see Table 2 of Ref.~\cite{pr87}), 
the authors had given the spin value of the 3631-keV level with parentheses, 
$I^\pi$= (9$^-$), because $I^\pi$= 7$^-$
could not be excluded. Since in the fusion-fission reactions the yrast states
are the most populated, the spin value of the 3631-keV level cannot be 
7$^-$ but is 9$^-$, without ambiguity. The same arguments hold for the 3909-keV level
($I^\pi$= 10$^+$) and 4687-keV level ($I^\pi$= 12$^+$), as the 727- and the 778-keV
$\gamma$ rays are stretched $E2$ transitions (see Table 2 of ref.~\cite{pr87}). 

The three first states of the high-energy part of the yrast structure of $^{138}$Ba are
proposed to have $I^\pi$= (11$^+$), (12$^+$), and (13$^+$) respectively, that are
consistent with their links to the 10$^+_1$ and 12$^+_1$ states, assuming that 
the spin values increase with excitation energy and that the low-energy 
transitions have a $M1$ character. Then the 4702-keV state is (11$^-$), since it is
populated by the (12$^+$) state at 5924 keV and it decays toward the 9$^-$ state at 
3631 keV and the 10$^+$ state at 3620 keV. Finally assuming that all the transitions 
located above the 6209-keV levels
have a $M1$ multipolarity, we assign $I^\pi$= (20$^+$) to the 9333-keV level.

\begin{figure}[!h]
\includegraphics[width=7cm]{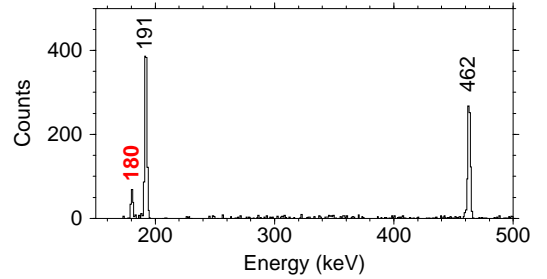}
\caption[]{(Color online) Spectrum of $\gamma$ rays which have been detected in the time interval
50~ns$-1\mu$s after the detection of two fragments by SAPhIR and in coincidence
with the 1435-keV line of $^{138}$Ba. The 180-keV transition is emitted by the isomeric 
state of $^{101}$Mo, partner of $^{138}$Ba in the C+U fusion-fission reaction.}
\label{saphir1435}
\end{figure}

We have looked for isomeric states in $^{138}$Ba by using the data registered with the
SAPhIR detector. Only one $\gamma$-ray cascade has been found to be delayed, that
decaying the known isomeric state at 2089~keV (see Fig.~\ref{schema_ba138}). The
spectrum of $\gamma$ rays which have been detected in the time interval
50~ns-1~$\mu$s after the detection of two fragments by SAPhIR and in coincidence
with the 1435-keV line is shown in Fig.~\ref{saphir1435}. In addition to the
expected 462- and 191-keV transitions belonging to the known delayed cascade, we 
observe a small peak at 180~keV. The latter is emitted by one
of the complementary fragments, $^{101}$Mo, which also exhibits an isomeric state
with T$_{1/2}$=95(15)~ns~\cite{re05}. Such a spectrum demonstrates the selectivity of 
an apparatus such as SAPhIR + Euroball when looking for the decay of isomeric states 
of the fission fragments. 

  
\subsubsection{Level scheme of $^{140}_{~58}$Ce\label{Ce140}}

$^{140}_{~58}$Ce is the heaviest $N=82$ isotone obtained in the fusion-fission
reactions used in the present work. Since it is in the light-$A$ tail of the Ce 
fragment distribution, its population is unfortunately low. Nevertheless
we have identified several new high-spin states located above 
the 13$^-$ state at 5102 keV, which was
the highest-spin level obtained in the ($\alpha$,2n) reaction~\cite{en86}.
Most of the new states have likely a negative parity, as they are only linked to the 13$^-$ state.
 
\begin{figure}[!h]
\includegraphics[width=8.5cm]{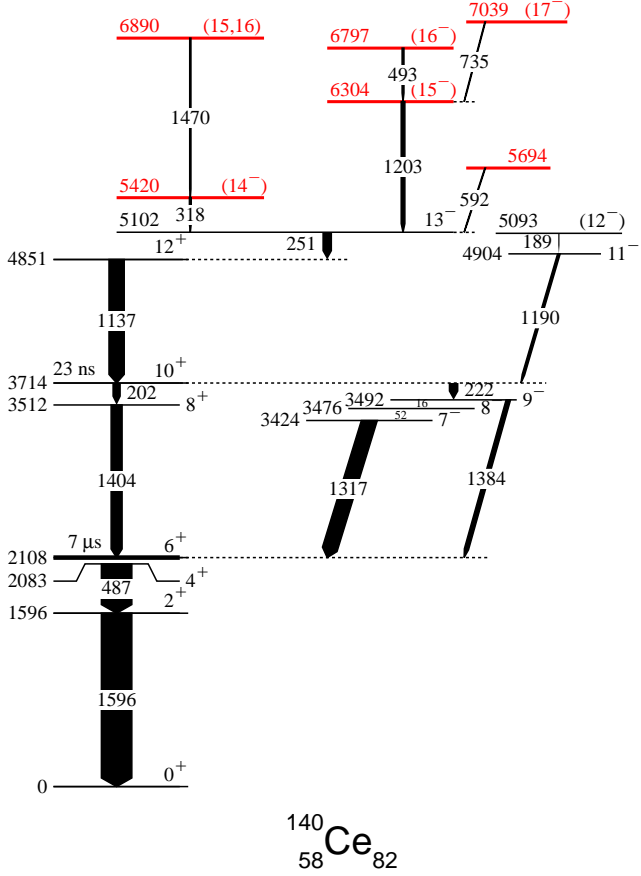}
\caption[]{(Color online) Level scheme of $^{140}$Ce established in this work.  
The colored states are new. The width of the arrows is proportional to the $\gamma$-ray intensity. The half-life values
of the 6$^+$ and 10$^+$ states are from Ref.~\cite{nndc}.
}
\label{schema_ce140}
\end{figure}

All the levels observed in this work are drawn in Fig.~\ref{schema_ce140}. Those lying below
5200 keV are the same as the yrast states previously obtained 
in the ($\alpha$,2n) reaction, their spin and parity values being determined from results
of angular distributions and linear polarizations~\cite{en86}.
The properties of the transitions observed in the present work are gathered in 
Table~\ref{gammas_Ce140}. Even though most of them were already known from the previous work,
they are repeated here for completeness.
\begin{table}[!h]
\caption{Properties of the transitions assigned to $^{140}$Ce observed in this work.}
\label{gammas_Ce140}
\begin{ruledtabular}
\begin{tabular}{rcccc}
$E_\gamma$\footnotemark[1](keV)& $I_\gamma$\footnotemark[1]$^,$\footnotemark[2]&  $I_i^\pi \rightarrow I_f^\pi$  &$E_i$&$E_f$  \\
\hline
16\footnotemark[3]	& -  & 9$^-$$\rightarrow$  8$^-$  & 3491.8 & 3476  \\  	
24.5\footnotemark[3]& -     & 6$^+$ $\rightarrow$  4$^+$  & 2107.6\footnotemark[4]& 2083  \\
52(1)	& -     & 8$^-$    $\rightarrow$  7$^-$   	& 3476	& 3424.4  \\  	
188.9(5) &4(2)& (12$^-$)    $\rightarrow$  11$^-$   	&5093.1&4904.2	\\
202.0(3) &28(6)& 10$^+$    $\rightarrow$  8$^+$  	&3713.9&3512.0	\\
222.0(3) &52(8)& 10$^+$    $\rightarrow$  9$^-$   	&3713.9&3491.8	\\
250.9(3) &33(7)& 13$^-$    $\rightarrow$  12$^+$   	&5101.6&4850.7	\\
318.0(4) &5.8(17)& (14$^-$)  $\rightarrow$  13$^-$   	&5419.6&5101.6	\\
487(1)\footnotemark[5]    &-       & 4$^+$    $\rightarrow$  2$^+$   	&2083  & 1596      \\
493.0(4)  &6.4(20)&  (16$^-$)    $\rightarrow$(15$^-$) 	&6797.2&6304.2	\\
592.3(5)  &4(2)  &     	$\rightarrow$  13$^-$ 		&5693.9&5101.6	\\
734.6(5)  &4(2)  &     (17$^-$)	$\rightarrow$(15$^-$)	&7038.8&6304.2	\\
1136.8(3) &51(8)& 12$^+$    $\rightarrow$  10$^+$   	&4850.7&3713.9	\\
1190.3(4) &8.8(25)& 11$^-$    $\rightarrow$  10$^+$   	&4904.2&3713.9	\\
1202.6(3) &13(3)&  (15$^-$)$\rightarrow$  13$^-$   	&6304.2&5101.6	\\
1316.8(3) &51(8)& 7$^-$    $\rightarrow$  6$^+$   	&3424.4&2107.6\footnotemark[4]	\\
1384.2(3) &14(4)& 9$^-$    $\rightarrow$  6$^+$   	&3491.8&2107.6\footnotemark[4]	\\
1404.4(3) &35(8)& 8$^+$    $\rightarrow$  6$^+$   	&3512.0&2107.6\footnotemark[4]	\\
1470.2(7) &4.3(17)  & (15,16)    $\rightarrow$(14$^-$) &6889.8&5419.6	\\
1596(1)\footnotemark[5]    &-        & 2$^+$    $\rightarrow$  0$^+$   	&1596   &0.0      \\	
\end{tabular}
\end{ruledtabular}
\footnotetext[1]{The number in parentheses is the error in the last digit.}
\footnotetext[2]{The relative intensities are normalized to the sum of the populations of 
the 6$^+_1$ isomeric
state, $I_\gamma(1405)+I_\gamma(1317) +I_\gamma(1385)= 100$.}
\footnotetext[3]{Not observed in the present work, from Ref.~\cite{en86}.}
\footnotetext[4]{Excitation energy from Ref.~\cite{en86}.}
\footnotetext[5]{Transition having a very weak intensity in this experiment because of
the long half-life of the 6$^+$ state.} 
\end{table}
For the spin assignments of the new levels, we have used the same arguments 
as those given in the previous sections. 
Lastly, we have looked for isomeric states in $^{140}$Ce by using the data registered with the
SAPhIR detector. Only the $\gamma$ rays decaying the two known isomeric states at
2108~keV and 3714~keV (see Fig.~\ref{schema_ce140}) are weakly observed.

\section{Discussion}\label{discuss}
The high-spin level schemes of the $N=82$ isotones with $Z > 50$ are expected to involve
only the three high-$j$ orbits, $\pi g_{7/2}$, $\pi d_{5/2}$, and $\pi h_{11/2}$. The
two former being very close to the Fermi level for $Z \le 58$, positive-parity
structures dominate the low-energy part of the spectra of the isotones studied in the
present work. In this section, we present general features of
their high-spin behaviors, particularly the maximum spin values which can be obtained
when all the proton pairs are broken; we discuss the breaking of one proton
pair for some simple configurations and we point out the main differences which may
occur in the odd-$Z$ isotones; we compare the results of 
shell-model calculations to the experimental levels, and lastly we show that the
excitation of the $N=82$ core is the only way to get the highest-spin
values of the lightest $N=82$ isotones.

\subsection{General features}\label{general}
Simple excitation modes are expected in the $N=82$ isotones lying just above the 
doubly-magic $^{132}$Sn because of the low number of available proton orbits. 
Being close in energy, the first two orbits, $\pi g_{7/2}$ and $\pi d_{5/2}$, can be
assumed to be filled together, thus the configurations are written as 
$(\pi g_{7/2} \pi d_{5/2})^n$, instead of $(\pi g_{7/2})^i (\pi d_{5/2})^j$, with $i+j=n$.
The maximum value of angular momentum, which can be achieved when
proton pairs are broken, depends on 
the total number of valence protons. For $^{134}_{52}$Te, the breaking of the proton pair
gives $I^\pi_{max}=6^+$, while one gets the very maximum value, $I^\pi=25/2^+$, 
for mid-occupation of the two orbits in $^{139}_{57}$La. Noteworthy is the fact that, due
to the presence of two different orbits, the same value of $I^\pi_{max}$ is obtained
for two different configurations of the even-$Z$ isotones, such as 6$^+$ for $(\pi g_{7/2})^2$ and 
$(\pi g_{7/2})^1 (\pi d_{5/2})^1$, or 10$^+$ for $(\pi g_{7/2})^3 (\pi d_{5/2})^1$ and 
$(\pi g_{7/2})^2 (\pi d_{5/2})^2$, meaning that we expect two 6$^+$ states or two 10$^+$
states in the corresponding nuclei.

The maximum values of angular momentum obtained for the complete alignment of the angular
momenta of the protons lying in the $\pi g_{7/2}$ and $\pi d_{5/2}$ orbits are given in
the first part of Table~\ref{spinmax}, for the $N=82$ isotones with $52 \le Z \le 58$.
These values are well lower than the ones which have been observed in the present work.
\begin{table}[!h]
\caption{Maximum values of angular momentum obtained from various configurations with 
several broken proton pairs, expected in the $N=82$ isotones with $52 \le Z \le 58$. 
The degeneracy, $d$, is the number of the configurations 
$(\pi g_{7/2})^i (\pi d_{5/2})^j$, $i+j=n$, leading to the same value of $I^\pi_{max}$.
}\label{spinmax}
\begin{ruledtabular}
\begin{tabular}{ccrr}
configuration&$I^\pi_{max}$&$d$&nucleus\\
		&&	&\\
\hline
$(\pi g_{7} \pi d_{5})^n$&&&\\
$n=2$			& 6$^+$	 	&2 & $^{134}$Te, $^{136}$Xe, $^{138}$Ba, $^{140}$Ce\\
$n=3$			& 17/2$^+$	&1 & $^{135}$I,  $^{137}$Cs, $^{139}$La\\
$n=4$			& 10$^+$	&2 & $^{136}$Xe, $^{138}$Ba, $^{140}$Ce\\
$n=5$			& 23/2$^+$	&1 & $^{137}$Cs, $^{139}$La\\
$n=6$			& 12$^+$	&2 & $^{138}$Ba, $^{140}$Ce\\
$n=7$			& 25/2$^+$	&1 & $^{139}$La \\
$(\pi g_{7} \pi d_{5})^n (\pi h_{11})^1$&&&\\
$n=1$ &9$^-$		&1 	&$^{134}$Te, $^{136}$Xe, $^{138}$Ba, $^{140}$Ce\\
$n=2$ &23/2$^-$	&2 	&$^{135}$I,  $^{137}$Cs, $^{139}$La\\
$n=3$ &14$^-$		&1 	&$^{136}$Xe, $^{138}$Ba, $^{140}$Ce\\
$n=4$ &31/2$^-$	&2 	&$^{137}$Cs, $^{139}$La\\
$n=5$ &17$^-$		&1 	&$^{138}$Ba, $^{140}$Ce\\
$n=6$ &35/2$^-$	&2 	&$^{139}$La \\
$n=7$ &18$^-$		&1 	&$^{140}$Ce \\
$(\pi g_{7} \pi d_{5})^n (\pi h_{11})^2$&&&\\
$n=1$ &27/2$^+$	&1 	&$^{135}$I,  $^{137}$Cs,   $^{139}$La\\
$n=2$ &16$^+$		&2 	&$^{136}$Xe, $^{138}$Ba  $^{140}$Ce\\
$n=3$ &37/2$^+$	&1 	&$^{137}$Cs,  $^{139}$La\\
$n=4$ &20$^+$		&2 	&$^{138}$Ba,  $^{140}$Ce\\
$n=5$ &43/2$^+$	&1 	&$^{139}$La\\
$n=6$ &22$^+$		&2 	&$^{140}$Ce  \\
\end{tabular}
\end{ruledtabular}
\end{table}
The promotion of one proton in the $\pi h_{11/2}$ orbit leads to higher values of angular
momenta as well as a change of parity (see the second part of Table~\ref{spinmax}). One
can remark that some negative-parity structures observed in $^{136}$Xe, $^{137}$Cs, and 
$^{138}$Ba agree with these quoted values of $I^\pi_{max}$. 
Finally the values of $I^\pi_{max}$ are increased by 10$\hbar$ when a proton pair
occupying the $\pi h_{11/2}$ orbit is broken (see the last part of Table~\ref{spinmax}). 
Nevertheless such a process is expected at high excitation energy in the nuclei of 
interest, since that orbit is located far from their Fermi levels.

Neutron excitation across the $N=82$ gap was proposed to be involved in  
the highest-spin part of the yrast band of $^{134}$Te, $^{135}$I, and 
$^{136}$Xe~\cite{zh96,da99}. The gain in angular momentum is 9$^+$ for the 
$(\nu h_{11/2})^{-1} (\nu f_{7/2})^{+1}$ excitation. While such an excitation is 
much more favorable for $Z=52-54$ than the breaking of a $\pi h_{11/2}$ pair mentionned 
above, it has not to be excluded for higher $Z$ values. Indeed two excited states of 
$^{137}$Cs lying at 4.35~MeV and 4.77~MeV were related to the excitation of the $N=82$
core~\cite{br99}. The clues of neutron excitation across the $N=82$ gap will be
presented in Sec.~\ref{SM} and discussed in Sec.~\ref{coreN82}, using the 
experimental results obtained in the present work. 

\subsection{Breaking of one proton pair}\label{onepair}

The {\it first} breaking of a proton pair in the $N=82$ isotones having $52 \le Z \le 58$ gives 
rise to different multiplets which depend on the total number of protons occupying the 
$\pi g_{7/2}$ and $\pi d_{5/2}$ orbits.

In the even-$Z$ isotones, we expect two simple configurations having a broken pair, 
$(\pi g_{7/2})^2$ and $(\pi g_{7/2})^1(\pi d_{5/2})^1$, 
the remaining protons lying as pairs in the $\pi g_{7/2}$ or/and $\pi d_{5/2}$ orbit 
according to the Pauli principle. Both configurations lead to $I^\pi_{max}=6^+$.
These two 6$^+$ states were identified in
\begin{figure}[!h]
\includegraphics[width=5.5cm]{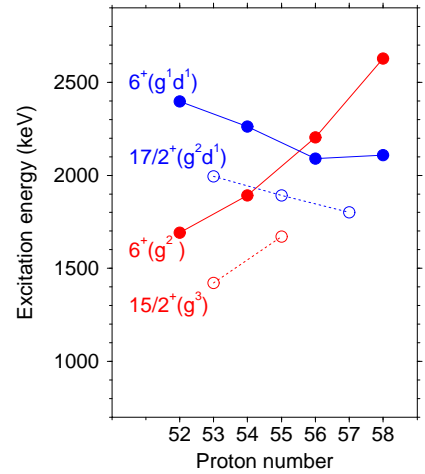}
\caption[]{(Color online) Evolution of the experimental two 6$^+$ states in the 
even-$Z$ isotones and of the 15/2$^+_1$ and 17/2$^+_1$ states in the odd-Z isotones
(data from this work and ref.~\cite{nndc}).}
\label{seniorites2et3}
\end{figure}
 $^{134}$Te, $^{136}$Xe, $^{138}$Ba, and $^{140}$Ce~\cite{zh96,da99,pr87,en86}. Their excitation energies 
evolve as a function of the position of the proton Fermi level within the two orbits, 
as shown in Fig.~\ref{seniorites2et3} (see the filled circles).  

\begin{figure}[!h]
\includegraphics[width=5.5cm]{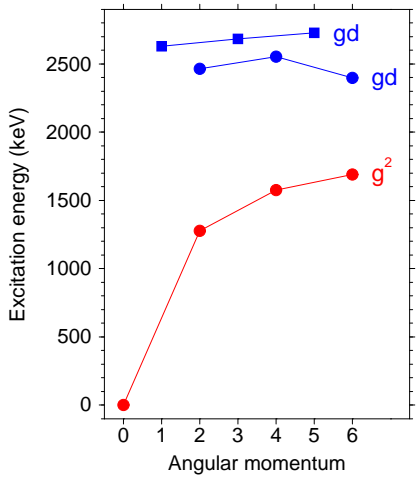}
\caption[]{(Color online) Experimental level scheme of $^{134}$Te showing the states issued from the 
$(\pi g_{7/2})^2$ and $(\pi g_{7/2})^1 (\pi d_{5/2})^1$ configurations~\cite{nndc}, with even  
values of angular momentum (circles) and odd ones (squares). 
}
\label{134Te}
\end{figure}

The case of the odd-$Z$ isotones is more delicate. If the odd proton is located in the 
$\pi g_{7/2}$ orbit, the {\it first} breaking of a $(\pi g_{7/2})^2$ pair leads to 
the $(\pi g_{7/2})^3$ configuration, with $I^\pi_{max}=15/2^+$. When the  odd proton is 
promoted to the $\pi d_{5/2}$ orbit, the multiplet of the 
$(\pi g_{7/2})^2(\pi d_{5/2})^1$ configuration
extends up to $I^\pi_{max}=17/2^+$. First, it is interesting to notice the
similar behaviors of the $I^\pi_{max}$ states of the odd-$Z$ and even-$Z$ isotones,  
the 15/2$^+$ level of  $(\pi g_{7/2})^3$ and the
6$^+$ level of $(\pi g_{7/2})^2$ on the one hand, the 17/2$^+$ level of 
$(\pi g_{7/2})^2(\pi d_{5/2})^1$ and the 6$^+$ level of  
$(\pi g_{7/2})^1(\pi d_{5/2})^1$ on the other hand 
(see Fig.~\ref{seniorites2et3}).
Secondly, it is worth noting that large changes in the structure of
the $(\pi g_{7/2})^2(\pi d_{5/2})^1$ multiplet are foreseen with $Z$, since the interaction between the three unpaired
protons depends on the total number of protons occupying the $\pi g_{7/2}$ orbit. Indeed
the two-body interaction evolves from an attractive particle-particle one (for low $Z$
value) to a repulsive particle-hole one (for high $Z$ value).
The evolution of the multiplets can be easily
predicted following the procedures described in Ref.~\cite{ta93}, which have been 
already used in similar cases~\cite{as11,as06,po04}. 
For that purpose, we only need
the values of the residual interactions in the $(\pi g_{7/2})^2$
and $(\pi g_{7/2})^1(\pi d_{5/2})^1$ configurations. They are extracted from the
multiplets of states identified in $^{134}_{~52}$Te (see Fig.~\ref{134Te}).

\begin{figure*}[!t]
\includegraphics[angle=-90,width=14cm]{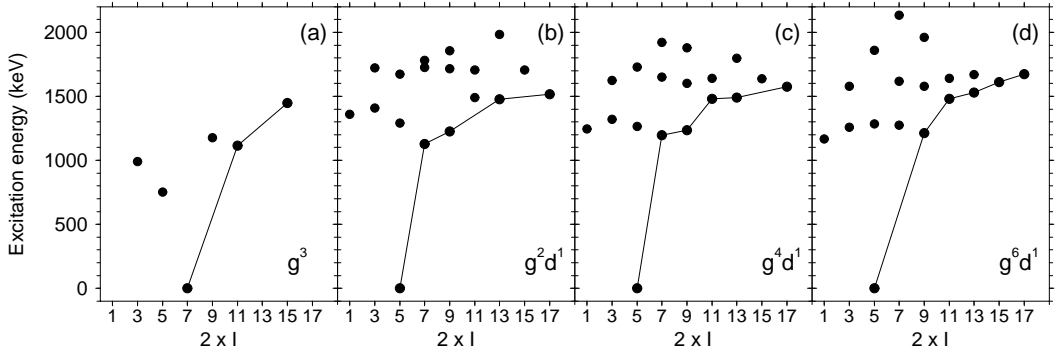}
\caption[]{States issued from four odd-Z configurations, $(\pi g_{7/2})^3$, 
$(\pi g_{7/2})^2 (\pi d_{5/2})^1$, $(\pi g_{7/2})^4 (\pi d_{5/2})^1$, and
$(\pi g_{7/2})^6 (\pi d_{5/2})^1$, calculated using the residual interactions
extracted from two multiplets of $^{134}$Te (see text). The yrast 
states are linked by a solid line.}
\label{config_g7d5}
\end{figure*}

The results of the computations of the $(\pi g_{7/2})^2(\pi d_{5/2})^1$ configuration 
are shown in Fig.~\ref{config_g7d5}, as compared to the results of the simple 
configuration, $(\pi g_{7/2})^3$, which is also active in the nuclei of interest.
The yrast
line of the latter comprises the 7/2$^+$, 11/2$^+$ and 
15/2$^+$ states [see Fig.~\ref{config_g7d5}(a)]. Noteworthy is the fact that,
due to the particle-hole symmetry, the
$(\pi g_{7/2})^{-3}$ configuration leads to the same states. On the other hand, three 
different sets are obtained for the $(\pi g_{7/2})^2(\pi d_{5/2})^1$ configuration, 
depending on the number of protons filling the $\pi g_{7/2}$ orbit:  
Fig.~\ref{config_g7d5}(b)
shows the results obtained for two particles in the $\pi g_{7/2}$ orbit, 
Fig.~\ref{config_g7d5}(c) for mid-occupation of the 
$\pi g_{7/2}$ orbit (i.e., for 4 particles) and Fig.~\ref{config_g7d5}(d) for two holes in the 
$\pi g_{7/2}$ orbit. It has to be noticed that the yrast
lines which extends from $I^\pi$=5/2$^+$ to 17/2$^+$ do not comprise exactly 
the same intermediate states. Thus the low-lying yrast states of the odd-$Z$ isotones 
with $N=82$ are expected to evolve with $Z$, as observed in the present work
(compare the bottoms of the $^{137}$Cs and $^{139}$La schemes,
Figs.~\ref{schema_cs137} and~\ref{schema_la139}). 
Moreover this very simple approach may explain why the 17/2$^+$ state can 
suddenly become an isomeric state, as measured in $^{139}$La, while no 
measurable half-life is found in the other odd-$Z$ isotones. 
Depending on the occupation rate of the $\pi g_{7/2}$ orbit, the 
15/2$^+$ state belongs to the yrast line or does not belong to it. In 
the first case, the $M1$ decay of the 17/2$^+$ state is fast, while in the
second case, the 17/2$^+$ state decays by emitting an $E2$ transition of low
energy [see, for instance, Fig.~\ref{config_g7d5}(c)], which leads to an
isomeric state. However we cannot make a one-to-one mapping between the four cases 
shown in Fig.~\ref{config_g7d5} and the odd-Z isotones, since the two
proton orbits are gradually filled together. Thus a shell-model approach with 
configuration mixings
has to be used to precisely discuss the yrast states of the $N=82$ isotones, 
this is presented in the next section. 

\subsection{Results of shell-model calculations}\label{SM}
Many years ago, a first shell-model (SM) analysis was carried out on the proton configurations of the
$N=82$ isotones lying above the doubly-magic $^{132}$Sn nucleus~\cite{wi69,wi71}. The 
two-body part of the SM
Hamiltonian was parametrized in terms of the modified surface delta interaction (MSDI).
The parameters of the interaction, as well as the single particle energies, were adjusted in
order to give the best fit of the experimental data known at that time. Many years later,
using the wealth of new data on the $N=82$ isotones, Wildenthal has modified a lot of 
the previously obtained values of the two-body matrix elements (TBME) in order to
fit as well as possible known excitations energies in nuclei from $^{133}$Sb to 
$^{154}$Hf~\cite{wi91}. Then Blomqvist has updated these TBME using new experimental
information in $^{133}$Sb~-~$^{138}$Ba: The new set of diagonal and non-diagonal
proton-proton matrix elements including the five orbits of the 50-82 major shell, are
given in Ref.~\cite{bl99}. Afterwards, these empirical effective interactions were used 
to describe the new yrast level scheme of $^{137}$Cs~\cite{br99}, the calculated energies
were found to be very close to the experimental ones.

In this section, we make a detailed comparison between the experimental information
obtained in the five $N=82$ isotones studied in the present work and the SM predictions 
using these empirical effective interactions~\cite{bl99}.   
The calculations were performed using the ANTOINE code~\cite{ca99}.
\subsubsection{The even-$Z$ isotones}\label{SM_even}

\begin{figure*}[!t]
\includegraphics[angle=-90,width=16cm]{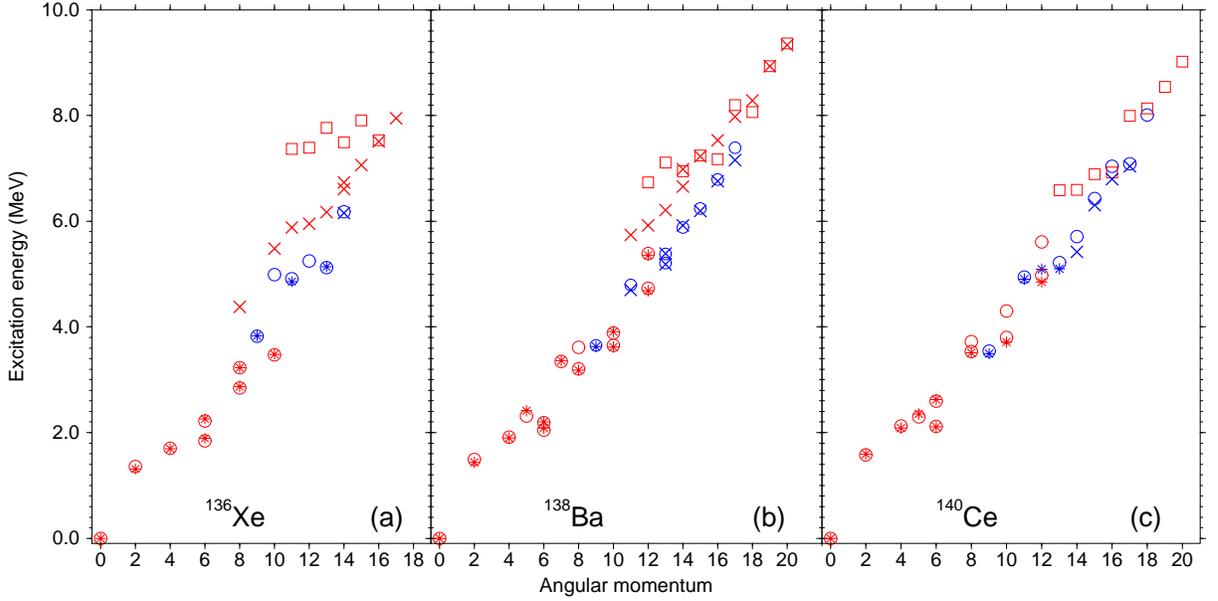}
\caption[]{(Color online) Excitation energy as a function of angular momentum of the
states of $^{136}$Xe (a), $^{138}$Ba (b), and $^{140}$Ce (c). 
The positive-parity states predicted by the SM calculations
are drawn with red circles when issued from the $(\pi g_{7/2}\pi d_{5/2})^n$
configurations and with red squares when issued from the 
$(\pi g_{7/2}\pi d_{5/2})^{n-2}(\pi h_{11/2})^2$ configurations. The negative-parity 
states predicted by the SM calculations are drawn with blue circles. The new 
experimental states identified in this work are drawn with crosses. The 
experimental states, which were used to fit the two-body matrix elements 
(see text), are drawn with asterisks. The red(blue) color is used for 
positive(negative) parity states.
}
\label{evenZ_exp_SM}
\end{figure*}

The results of the SM calculations of the yrast states of $^{136}$Xe, $^{138}$Ba, and
$^{140}$Ce are drawn in Fig.~\ref{evenZ_exp_SM}, where they are compared to the experimental
results. 
The positive-parity states up to the 10$^+_1$ level in $^{136}$Xe  
and to the 12$^+_1$ level in $^{138}$Ba and $^{140}$Ce are obviously well reproduced 
(see the red asterisks and circles),
since they took part of the data set used to fit the two-body matrix 
elements~\cite{bl99}. For the negative-parity states (drawn in blue), the SM 
results show that the yrast structures strongly depend on the available
configurations. While $^{136}$Xe only displays 
the $14^-\rightarrow 13^-\rightarrow 11^-\rightarrow 9^-$ sequence from the 
$(\pi g_{7/2}\pi d_{5/2})^3(\pi h_{11/2})^1$ configuration, the addition of
the $(\pi g_{7/2}\pi d_{5/2})^5(\pi h_{11/2})^1$ configuration in $^{138}$Ba and 
$^{140}$Ce leads to almost regular structures extending up to I$^\pi=17^-$. This is 
in good agreement with the new levels observed in $^{138}$Ba and $^{140}$Ce [see the
blue crosses in Fig.~\ref{evenZ_exp_SM}(b) and (c)]. 
More precisely, two 13$^-$ states are predicted close in energy in $^{138}$Ba, as 
measured experimentally. Their main configurations explain their different decays. 
The  13$^-_1$ and 11$^-_1$ states have the same configuration, 
$(\pi g_{7/2})^5(\pi h_{11/2})^1$, while the 13$^-_2$ state has another one, 
$(\pi g_{7/2})^4(\pi d_{5/2})^1(\pi h_{11/2})^1$. Moreover, the latter configuration
is the one of the 14$^-_1$ state, explaining why this state does not decay to the 13$^-_1$ 
state but is only linked to the 13$^-_2$ state.

As mentioned before (see Table~\ref{spinmax}), when restricting excitations to the
proton valence space, the high values of the angular
momenta need to promote a proton pair into the $\pi h_{11/2}$ orbit and to break 
that pair. In that way and 
by breaking another pair in the $\pi g_{7/2}$ and $\pi d_{5/2}$ orbits, states 
with I$^\pi$ values up to 16$^+$ are obtained. The multiplet (12$^+$ $\le$ I $\le$ 16$^+$) 
is predicted almost degenerated
in energy in the three isotones [see the red squares in Fig.~\ref{evenZ_exp_SM}]. 
In $^{138}$Ba and $^{140}$Ce, such a structure extends up to higher 
spin values, as more pairs of protons lying in the  $\pi g_{7/2}$ and $\pi d_{5/2}$ 
orbits can be broken. The resulting
levels (16$^+$ $\le$ I $\le$ 20$^+$) are no longer degenerated, being spread over an interval
of $\sim$~2 MeV [see the upper red squares in Figs.~\ref{evenZ_exp_SM}(b) and (c)]. 
The new structure of $^{136}$Xe, which extends from the 5481-keV state to the 
7946-keV one [see the red crosses in Fig.~\ref{evenZ_exp_SM}(a)], is too low in 
energy to come from the $(\pi g_{7/2})^2(\pi h_{11/2})^2$ configuration. It will be
discussed in Sec.~\ref{coreN82} in terms of neutron excitation across the $N=82$ gap, as
already done for the highest-spin states observed in $^{134}$Te~\cite{zh96}. The behavior 
of $^{138}$Ba is
more complex. The SM results fit reasonably well the high part of the experimental sequence 
(from the (14$^+_2$) state to the (20$^+$) one) but not its low part (from the (11$^+$) 
state to the (14$^+_1$) one). Thus the latter is likely issued from the neutron 
excitation across the $N=82$ gap, while the configuration of the former is likely 
$(\pi g_{7/2}\pi d_{5/2})^4(\pi h_{11/2})^2$. 

For $^{140}$Ce, because of its low
population in the reactions used in the present work, no cascade of low-energy 
transitions lying in the top of its level scheme could be observed, thus its
highest-spin states with positive parity remain unobserved.

\subsubsection{The odd-$Z$ isotones}\label{SM_odd}

\begin{figure}[!h]
\includegraphics[width=7.5cm]{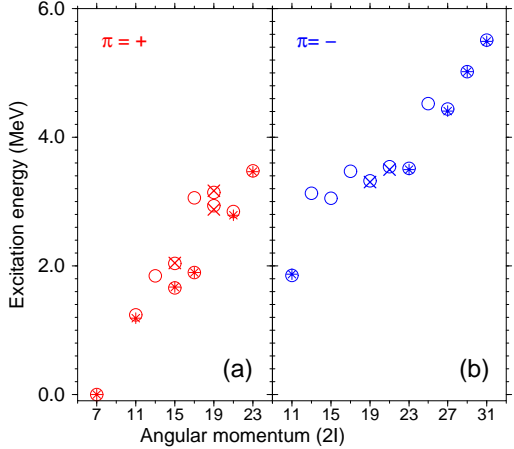}
\caption[]{(Color online) Excitation energy as a function of angular momentum of the
states of $^{137}$Cs. The states predicted by the SM calculations for the 
$(\pi g_{7/2}\pi d_{5/2})^5$ (a) and $(\pi g_{7/2}\pi d_{5/2})^4(\pi h_{11/2})^1$ configuration (b)
are drawn with circles. The new experimental states identified in this work
are drawn with crosses. The already known experimental states are drawn with asterisks, 
a few of them were used to fit the two-body matrix elements (see text).
}
\label{137Cs_exp_SM}
\end{figure}

When Blomqvist had updated the TBME values~\cite{bl99}, the states of $^{137}$Cs  having 
I$^\pi \geq$ 11/2 were not yet identified, thus they were not considered in
the fit. A few time later, the newly-observed levels of $^{137}$Cs were compared to the 
theoretical results showing a very good agreement~\cite{br99}. Results of the calculations of the 
yrast states, as well as some of them close to the yrast line, are shown in Fig.~\ref{137Cs_exp_SM} 
(see the circles), in comparison with the experimental results (drawn with asterisks and crosses). 
The theoretical yrast line displays several 
irregularities. For instance, the two 19/2$^+$ states are located above the first 21/2$^+$ state, 
thus the latter decays directly toward the 17/2$^+$ state. As well, the
13/2$^+_1$ state is located higher in energy than the 15/2$^+_1$ one, thus the latter  
decays toward the 11/2$^+$ state.  

The negative-parity states behave
similarly. The 25/2$^-_1$ state is located higher in energy than the 27/2$^-_1$ one, 
thus the latter decays toward the 23/2$^-$ state, as observed experimentally. The predicted behavior 
of the first states is very peculiar, showing a quasi-degeneracy of the six states with 
I$^\pi$ = 13/2$^-$ - 23/2$^-$. This explains why the three observed states 
(19/2$^-$, 21/2$^-$, and 23/2$^-$) only decay to the positive-parity states.

\begin{figure}[!h]
\includegraphics[width=4.2cm]{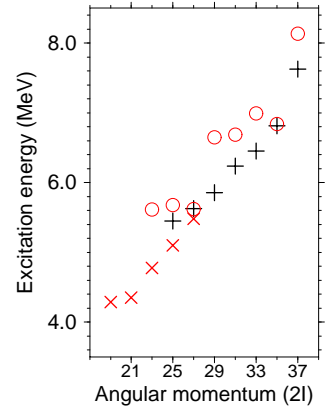}
\caption[]{(Color online) High-energy level scheme of $^{137}$Cs. States of structure C
are drawn with $\times$ and those of structure D with $+$.  
The SM states from the 
$(\pi g_{7/2}\pi d_{5/2})^3(\pi h_{11/2})^2$ configuration are drawn with circles. 
}
\label{137Cs_exp_SM_highpart}
\end{figure}

As discussed in Sec.~\ref{general}, a proton pair has to be promoted to the 
$\pi h_{11/2}$ orbit in 
order to describe the higher values of angular momentum (see the bottom part of Table~\ref{spinmax}). 
The calculated states of $^{137}$Cs having the $(\pi g_{7/2}\pi d_{5/2})^3(\pi h_{11/2})^2$ configuration are 
shown in Fig.~\ref{137Cs_exp_SM_highpart} (see the circles). 
The yrast line is irregular, several states being predicted very close in energy. Nevertheless,
this line is not far from the states of Structure D (see the $+$ symbols in 
Fig.~\ref{137Cs_exp_SM_highpart}). Thus
the main configuration of the states of Structure D is likely 
$(\pi g_{7/2}\pi d_{5/2})^3(\pi h_{11/2})^2$. 
On the other hand, states of Structure C are too low in energy to be explained in terms of proton
excitations (see the $\times$ symbols in 
Fig.~\ref{137Cs_exp_SM_highpart}). As already discussed in Sec.~\ref{SM_even} for $^{136}$Xe and $^{138}$Ba, neutron
excitation across the $N=82$ gap has to be considered (cf. the next section, Sec.~\ref{coreN82}).

\begin{figure}[!h]
\includegraphics[width=8cm]{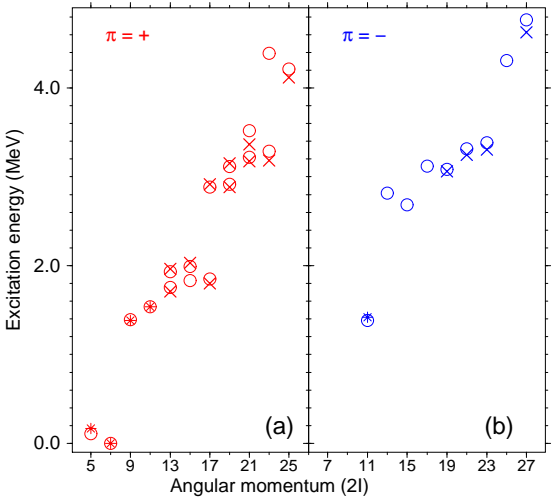}
\caption[]{(Color online) Excitation energy as a function of angular momentum of the
states of $^{139}$La. The states predicted by the SM calculations
are drawn with circles. The new experimental states identified in this work
are drawn with crosses. The experimental states, which were used to fit the
two-body matrix elements (see text), are drawn with asterisks. 
}
\label{139La_exp_SM}
\end{figure}

Results of calculation performed for $^{139}$La are shown in 
Fig.~\ref{139La_exp_SM} (see the circles), where we have drawn one or two
states for each spin value. First of all, it is important to notice that the
positive-parity yrast line is different from the results obtained in $^{137}$Cs. Indeed 
several states of $^{139}$La are found at lower energy and can now take part to the 
yrast decay, such as the 9/2$^+_1$, 13/2$^+_1$, and  19/2$^+_1$ states. This change is 
related to the discussion done in Sec.~\ref{onepair}.  It is worth noting that the 15/2$^+_1$ 
state is predicted to lie 20~keV below the 17/2$^+_1$ state, thus the latter would not exhibit
any delayed decay. This is at variance with the experimental results showing an isomeric state
with T$_{1/2}$~=~315~ns. We surmise that the energy of the 15/2$^+_1$ state is predicted
slightly too low. Such a discrepancy is in agreement with the observed deviations between 
the experimental and calculated energies, see for instance Table III of Ref.~\cite{bl99} for
the light-$A$ isotones and Table~\ref{E_exp_SM_139Ba} (this work) for $^{139}$La.  

\begin{table}[!h]
\caption{Comparison of experimental and calculated energies of the yrast states of $^{139}$La. The TBME
values used in the SM calculations are those of Ref.~\cite{bl99}.}\label{E_exp_SM_139Ba}
\begin{ruledtabular}
\begin{tabular}{cccc}
I$^\pi$ & E$_{exp}$ & E$_{SM}$ 	& E$_{exp}$-E$_{SM}$\\
	& (keV)	    & (keV)	& (keV)\\
\hline	
7/2$^+$& 	0& 	0&	0\\
5/2$^+$& 	166&	109&	+57\\ 
9/2$^+$&	1381&  	1394&	-13\\
11/2$^-$&	1420&  	1382&	+38\\
11/2$^+$& 	1537& 	1538&	-1\\
13/2$^+_1$& 	1711& 	1756&	-45\\
15/2$^+_1$& 	-& 	1832&	\\
17/2$^+$& 	1800& 	1852&	-52\\
13/2$^+_2$& 	1962& 	1933&	+29\\
15/2$^+_2$& 	2032& 	1991&	+41\\
19/2$^+_1$& 	2885& 	2915&	-30\\
17/2$^+$& 	2917& 	2882&	+35\\
19/2$^-$& 	3060& 	3085&	-25\\
19/2$^+_2$& 	3150& 	3115&	+35\\
21/2$^+_1$& 	3175& 	3218&	-43\\
23/2$^+$& 	3184& 	3287&	-103\\
23/2$^-$& 	3305& 	3384&	-79\\
21/2$^+_2$& 	3364& 	3518&	-154\\
25/2$^+$& 	4116& 	4214&	-98\\
27/2$^-$& 	4627& 	4765&	-138\\
\end{tabular}
\end{ruledtabular}
\end{table}

All the experimental states observed below 4116 keV have a theoretical counterpart 
(see the stars and the crosses in Fig.~\ref{139La_exp_SM}).  
For the 15/2$^+$
states, only one experimental level is proposed at 2032~keV, which is close to the prediction
of the 15/2$^+_2$ state. The fact that the 15/2$^+_1$ state was not populated in our
experiment could be explained by its configuration,$(\pi g_{7/2})^6(\pi d_{5/2})^1$, which
is different from the one of the states lying above it: The 17/2$^+_2$, 19/2$^+_1$, 
and 19/2$^+_2$ states have the same configuration, $(\pi g_{7/2})^5(\pi d_{5/2})^2$.
It is important to note that the experimental 21/2$^+_2$, 23/2$^+$ and 25/2$^+$ states 
are located at lower energies than the calculated ones, the deviations being $\sim$~100~keV 
(Table~\ref{E_exp_SM_139Ba}). This could be related to interactions between the states
due to proton excitations among the $\pi g_{7/2}$ and $\pi d_{5/2}$ orbits (belonging to the
valence space used in the SM calculations) and the states of Structure C 
(see Fig.~\ref{schema_la139}).

Finally, the states of Structure C would be likely due to the breaking of the neutron core, as
the corresponding states of $^{137}$Cs. Nevertheless since the high-spin states having a broken 
proton pair
in the $\pi h_{11/2}$ orbit are predicted at lower energy than in $^{137}$Cs 
(about 600~keV less), we expect large mixings of these two excitation modes in $^{139}$La.  

\subsubsection{Conclusion}
This set of empirical effective interactions within the proton valence
space including all the orbits of the 50-82 major shell describes well the 
yrast states of the five isotones studied in the present work, both the 
even-$Z$ and the odd-$Z$ ones. The deviations between experimental and calculated 
energies are mostly below 50~keV, that is well better than the values obtained for SM
calculations using effective interactions derived from nucleon-nucleon potential, such as
done for one prediction of the yrast states of $^{137}$Cs (see Table II of Ref.~\cite{li07}). The 
latter approach is broader since it can be used in all the valence spaces, thus it would 
be interesting to compare the numerical values of the two sets in order to find whether
some particular TBME are at the origin of the better description and to understand why
they would be unsatisfactorily calculated from nucleon-nucleon potential.

\subsection{Neutron-core excitation}\label{coreN82}
\begin{figure}[!h]
\includegraphics[width=8cm]{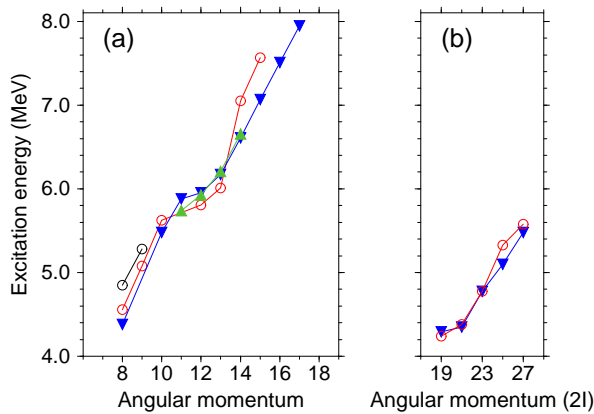}
\caption[]{(Color online) Excitation energy as a function of angular momentum 
of the states proposed to be due to neutron excitation across the $N=82$ gap, i.e.  
having the main configuration,  
$(\pi g_{7/2}\pi d_{5/2})^n \otimes (\nu h_{11/2})^{-1}(\nu f_{7/2})^{+1}$. (a)  
$n=0$ for $^{132}$Sn (black circles), $n=2$ for $^{134}$Te (red circles), 
$n=4$ for $^{136}$Xe (blue triangles), and $n=6$ for $^{138}$Ba (green triangles). 
(b) $n=3$ for $^{135}$I (red circles) and 
$n=5$ for $^{137}$Cs (blue triangles) (data from Ref.~\cite{nndc} and this work).
}
\label{coeurbrise}
\end{figure}

The neutron excitation of the doubly-magic $^{132}$Sn nucleus gives rise to its
first positive-parity excited states: The multiplet from the 
$(\nu h_{11/2})^{-1}(\nu f_{7/2})^{+1}$ configuration was measured in the 4-5 MeV 
energy range, with spin values extending from 2$^+$ to 9$^+$~\cite{fo95,ba01}. 
In $^{134}$Te, states with higher spin values are obtained as this first neutron-core 
excitation can be coupled to the breaking of the proton pair~\cite{zh96} 
[see the red circles in Fig.~\ref{coeurbrise}(a)].  
The states of $^{136}$Xe, drawn with crosses in Fig.~\ref{evenZ_exp_SM}(a), 
extend from
4379~keV [I$^\pi$ = (8$^+$)] to 7946~keV [I$^\pi$ = (17$^+$)]. They fit well with those due 
to the neutron-core breaking of $^{134}$Te [compare the blue triangles and the red circles in 
Fig.~\ref{coeurbrise}(a)]. Finally, the states of  $^{138}$Ba extending from 5740~keV 
[I$^\pi$ = (11$^+$)] to 6656~keV [I$^\pi$ = (14$^+$)] are in good agreement with those of
the lighter isotones [see the green triangles in Fig.~\ref{coeurbrise}(a)]. The closeness
of the states of $^{134}$Te, $^{136}$Xe, and $^{138}$Ba involving the neutron-core 
excitation is due to the stability of the gap in energy between the $\nu h_{11/2}$ and the
$\nu f_{7/2}$ orbits, as shown in Fig.~38 of Ref.~\cite{so08}. 

For the breaking of the neutron core in the odd-Z isotones, we compare in 
Fig.~\ref{coeurbrise}(b) the states of $^{135}$I lying between 4241~keV and 
5576~keV~\cite{zh96} and those of Structure C of $^{137}$Cs, both of them being not 
described by SM calculations within the proton space (see Ref.~\cite{zh96} and 
Sec.~\ref{SM_odd}). They show the same behavior and are in good
agreement with the corresponding states of the even-$Z$ isotones.

Even though the configurations of the highest-spin states of $^{137}$Cs and 
$^{138}$Ba observed in the present work can be $(\pi g_{7/2}\pi d_{5/2})^3(\pi h_{11/2})^2$ and 
$(\pi g_{7/2}\pi d_{5/2})^4(\pi h_{11/2})^2$, respectively (see Sects.~\ref{SM_odd} 
and~\ref{SM_even}), we cannot exclude other configurations involving the breaking of
the neutron core, such as 
$(\pi g_{7/2}\pi d_{5/2})^5(\nu h_{11/2})^{-1}(\nu f_{7/2})^{+1}$ and
$(\pi g_{7/2}\pi d_{5/2})^6(\nu h_{11/2})^{-1}(\nu f_{7/2})^{+1}$, respectively. 

\section{Summary}
Five $N=82$ isotones have been produced as fission fragments in two fusion reactions, 
$^{12}$C + $^{238}$U at 90~MeV and $^{18}$O + $^{208}$Pb at 85~MeV. Their high-spin level
schemes have been built by analyzing the triple $\gamma$-ray coincidence data obtained with
the Euroball array. In order to identify the new transitions of $^{139}$La, we have used their
coincidences with the lines emitted by its complementary fragments. Moreover, the use of the
fission-fragment detector, SAPhIR, has allowed us to identify a new isomeric state 
in $^{139}$La, at 1800-keV excitation energy. Spin and parity
values of several excited states of $^{136}$Xe and $^{137}$Cs could be assigned from 
the results of $\gamma-\gamma$ angular correlations. 
All the states observed in these $N=82$ isotones have been compared to results of shell
model calculations performed in the 50-82 proton valence space and using empirical 
effective interactions which had been fitted on previous experimental data. Most of the yrast
states of the five $N=82$ isotones are very well described by this approach. In addition, 
the excitation of the $N=82$ core gives rise to several structures which have been clearly
identified in $^{136}$Xe, $^{137}$Cs, and $^{138}$Ba. 

\begin{acknowledgments}
The Euroball project was a collaboration among France, the 
United Kingdom, Germany, Italy, Denmark and Sweden. 
The first experiment has been performed under 
U.E. contract (ERB FHGECT 980 110) at Legnaro. 
The second experiment has 
been supported in part by the EU under contract HPRI-CT-1999-00078 (EUROVIV). 
We thank many colleagues for their
active participation in the experiments, Drs. A.~Bogachev, A.~Buta, J.L.~Durell, 
Th.~Ethvignot, F.~Khalfalla, I.~ Piqueras, A.A.~Roach, A.G.~Smith and B.J.~Varley. 
We thank the crews of the Vivitron. 
We are very indebted to M.-A. Saettle
for preparing the Pb target, P. Bednarczyk, J. Devin, J.-M. Gallone, 
P. M\'edina, and D. Vintache for their help during the experiment. 
\end{acknowledgments}

\end{document}